\documentclass[conference]{IEEEtran}

\usepackage{lscape}
\usepackage{multirow}
\usepackage{graphicx}
\usepackage{booktabs}

\usepackage{algorithm}
\usepackage{algpseudocode}
\usepackage{amsmath}
\usepackage{graphicx}

\usepackage[numbers,sort&compress]{natbib}

\usepackage{bm}

\usepackage{xcolor}
\usepackage{xspace}

\usepackage{colortbl}
\usepackage{booktabs} 

\usepackage{subcaption}

\PassOptionsToPackage{breaklinks = true,unicode = true,urlcolor = blue,colorlinks = true,citecolor = red,linkcolor = red}{hyperref}
\usepackage{hyperref}

\usepackage{multirow}
\usepackage{array}
\newcolumntype{P}[1]{>{\centering\arraybackslash}p{#1}} 

\usepackage{enumitem}

\usepackage{multicol}
\usepackage{multirow}

\usepackage{tabularx}
\usepackage{balance}
\usepackage{epstopdf}

\usepackage{amsfonts}
\usepackage{amsthm}

\usepackage[nameinlink,capitalize]{cleveref} 

\usepackage[font=small]{caption}

\newcommand{\paragraphb}[1]{\vspace{0.75ex}\noindent{\bf #1.} }

\newcommand{\paragraphq}[1]{\vspace{0.75ex}\noindent{\bf #1} }

\newcommand{\ignore}[1]{}

\definecolor{orange}{RGB}{255,127,80}
\definecolor{darkgreen}{RGB}{50,127,0}
\definecolor{Blue}{RGB}{0,0,255}

\usepackage{comment}
\usepackage{amsmath}
\DeclareMathOperator*{\argmax}{arg\,max}
\DeclareMathOperator*{\argmin}{arg\,min}

\newcommand{\sgd}{SGD}
\newcommand{\dpsgd}{DP-SGD}

\newcommand{\mnist}{MNIST}
\newcommand{\cifar}{CIFAR-10}
\newcommand{\fmnist}{Fashion-MNIST}

\newcommand{\svhn}{SVHN}
\newcommand{\adult}{Adult}
\newcommand{\purchase}{Purchase-100}
\newcommand{\breast}{Breast Cancer}
\newcommand{\spam}{SMS Spam Collection}
\newcommand{\medical}{Medical Cost}

\newcommand{\randomweights}{RWT}

\newcommand{\genticrandom}{MGRS}
\newcommand{\randomsearch}{RS}

\newcommand{\pr}[1]{\ensuremath{\mathrm{Pr}(#1)}} 

\newcommand{\mc}[1]{\ensuremath{\mathcal{#1}}}

\newcommand{\inspace}{\ensuremath{X}}
\newcommand{\labspace}{\ensuremath{Y}}
\newcommand{\outspace}{\labspace}

\newcommand{\dpalg}{\ensuremath{F}}

\newcommand{\train}{\ensuremath{{\textsf{train\_model}()}}}

\newcommand{\dsdist}{\ensuremath{\mc{D}}}
\newcommand{\ds}{\ensuremath{{\bf{d}}}}
\newcommand{\datapoint}{\ensuremath{{\bf x}}}
\newcommand{\x}{\datapoint}

\newcommand{\lab}{\ensuremath{y}}
\newcommand{\y}{\lab}

\newcommand{\model}{\ensuremath{h}}
\newcommand{\modelspace}{\ensuremath{H}}

\newcommand{\params}{\ensuremath{{\bm{\theta}}}}
\newcommand{\paramsspace}{\ensuremath{{\bm{\Theta}}}}

\newcommand{\loss}{\ensuremath{l}}

\newcommand{\suc}{\ensuremath{{\rm suc}}}
\newcommand{\merit}{\ensuremath{{\rm merit}}}
\newcommand{\corr}{\ensuremath{{\rm corr}}}

\newcommand{\err}[1]{\ensuremath{{\rm err}(#1)}}
\newcommand{\priverr}[2]{\ensuremath{{\rm err}_{#1}(#2)}}

\newcommand{\cost}[1]{\ensuremath{{\rm cost}}(#1)}

\newtheorem{definition}{\bf Definition}
\newtheorem{corollary}{\bf Corollary}
\newtheorem{lemma}{\bf Lemma}

\begin{document}

\title{On the Importance of Architecture and Feature Selection in Differentially Private Machine Learning} %

\author{
    \IEEEauthorblockN{Wenxuan Bao, Luke A. Bauer, and Vincent Bindschaedler}
    \IEEEauthorblockA{University of Florida
    \\\{wenxuanbao, lukedrebauer, vbindschaedler\}@ufl.edu}
}

\maketitle

\thispagestyle{plain}
\pagestyle{plain}

\begin{abstract}
  We study a pitfall in the typical workflow for differentially private machine learning. The use of differentially private learning algorithms in a ``drop-in'' fashion --- without accounting for the impact of differential privacy (DP) noise when choosing what feature engineering operations to use, what features to select, or what neural network architecture to use --- yields overly complex and poorly performing models. In other words, by anticipating the impact of DP noise, a simpler and more accurate alternative model could have been trained for the {\em same privacy guarantee}. We systematically study this phenomenon through theory and experiments. On the theory front, we provide an explanatory framework and prove that the phenomenon arises naturally from the addition of noise to satisfy differential privacy. On the experimental front, we demonstrate how the phenomenon manifests in practice using various datasets, types of models, tasks, and neural network architectures. We also analyze the factors that contribute to the problem and distill our experimental insights into concrete takeaways that practitioners can follow when training models with differential privacy. Finally, we propose {\em privacy-aware} algorithms for feature selection and neural network architecture search. We analyze their differential privacy properties and evaluate them empirically.
\end{abstract}

\section{Introduction}\label{sec:intro}
Over the past decade, researchers have developed versions of traditional learning algorithms to train machine learning models in a way that provably satisfies differential privacy~\cite{dwork2006calibrating,dwork2014algorithmic}. This includes output and objective perturbation~\cite{chaudhuri2008privacy,chaudhuri2011differentially}, the functional mechanism~\cite{zhang2012functional}, mechanisms for decision trees~\cite{jagannathan2009practical,friedman2010data}, and more recently Differentially Private Stochastic Gradient Descent (\dpsgd)~\cite{abadi2016deep} or its variants~\cite{nasr2020improving,papernot2020tempered,davody2020robust,zhou2020bypassing} that can be used as a replacement of the traditional Stochastic Gradient Descent (SGD)~\cite{gower2019sgd} algorithm.

These differentially private learning algorithms are significant for the prospect of training machine learning models while preserving privacy. However, as we show in this paper, using these algorithms in a ``drop-in'' fashion in the workflow of ML engineers and practitioners is a subtle but serious pitfall. We show through experiments that the results of this pitfall are disastrous: models obtained this way consistently exhibit poor performance where significantly better performing models (with fewer parameters) exist and {\em could have been trained instead} for the exact same privacy guarantee. The reason is that simply replacing a traditional learning algorithm with a differentially private version of it, ignores the effect of noise added to achieve differential privacy and does not allow ML engineers and practitioners to make choices that account for it. As we discover in experiments, choices about feature engineering, feature selection, architecture selection, and hyperparameter tuning all impact differential privacy.

\begin{figure}[!t]
\centering
    \includegraphics[width=0.95\linewidth]{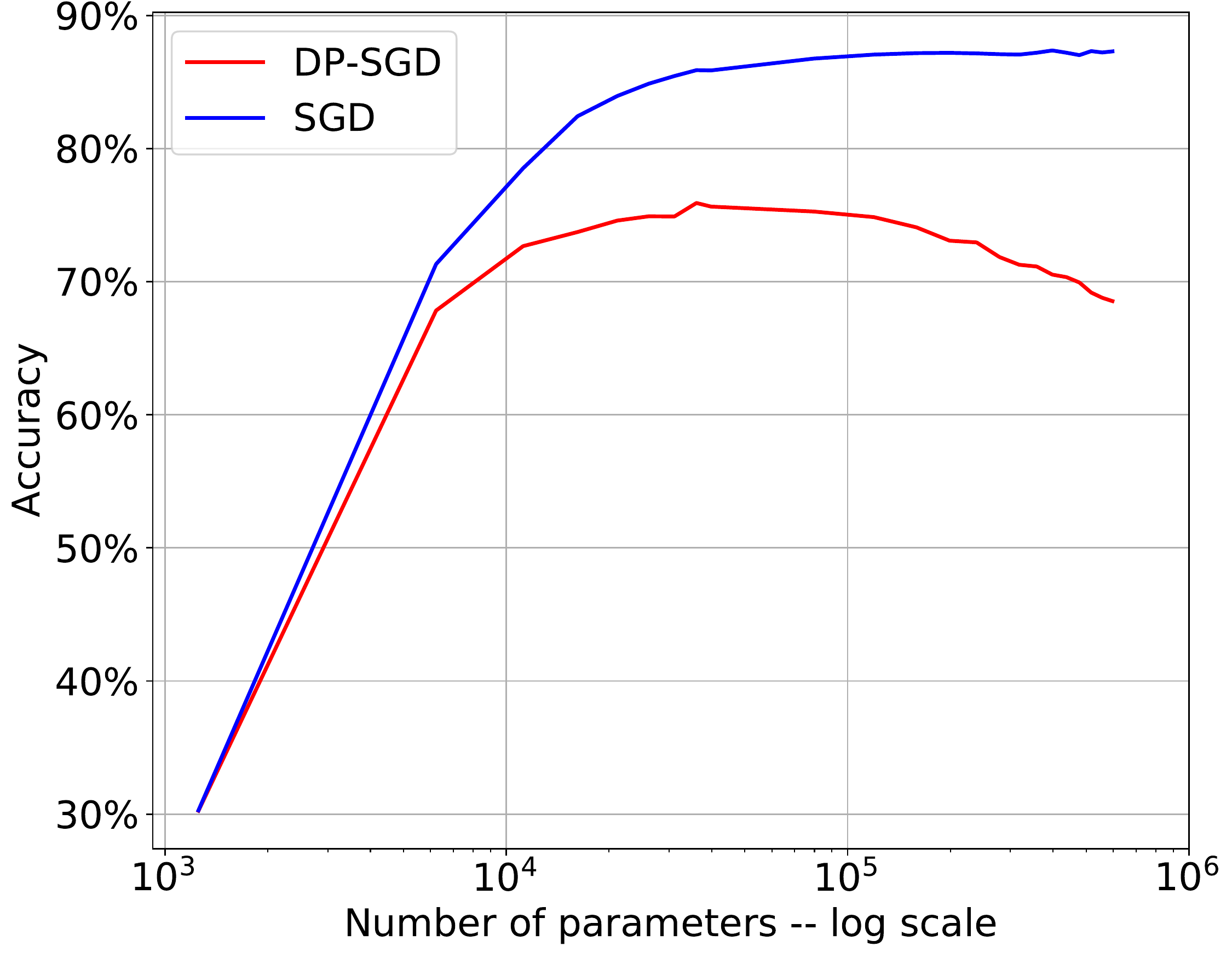}
    \caption{Test accuracy as a function of complexity for models trained with and without differential privacy (\sgd{} versus \dpsgd{}).}
    \label{fig:intro-complex}
    \vspace{-6pt}
\end{figure}

\cref{fig:intro-complex} shows the accuracy of increasing neural networks architectural complexity (measured in number of trainable parameters) when the learning algorithm is \sgd{} (blue curve) and \dpsgd{} (red curve). The gap between the blue and red curves can be thought of as the {\em cost of (achieving) privacy}. But this gap is not uniform with respect to complexity. Indeed, if a machine learning practitioner wants to train a highly accurate model with \sgd{} then the most suitable architecture is also the most complex one. By contrast, if the goal is to train a differentially private model (with \dpsgd) then an architecture with a middle-of-the-range complexity will provide the model with the highest accuracy. In plain words: {\bf if past wisdom about neural network architecture design is relied upon and (or) choices made following a typical machine learning workflow do not account for the {\em cost of achieving differential privacy}, a worse-than-necessary model will be trained}.

In this paper, we systematically study this phenomenon with respect to feature selection and neural network architecture search. We provide an explanatory theory, which conjectures that the effect of noise added to achieve differential privacy is such that the distortion is greater for more complex models. This leads to a tradeoff between the benefit of increased complexity (in terms of predictive power) and its cost (as a result of differentially private training). We prove that the effect of differential privacy conforms to what is predicted by our theory in the case of linear models where differential privacy is achieved by adding Gaussian noise to the model parameters. We also identify a measurable quantity that practitioners can use to avoid the pitfall: the {\em crossover epsilon ($\varepsilon^\times$)}, which is the threshold differential privacy budget under which the problem manifests. That is, if the target privacy budget is smaller than $\varepsilon^\times$, a worse-than-necessary model may be trained.

To validate our theory and assess the pervasiveness and impact of the pitfall in practice, we perform experiments using several datasets, various types of models, tasks, and neural network architectures. 

\paragraphb{Feature selection}
The use of a small subset of carefully chosen features (or in some cases a single feature) provides a better $\varepsilon$-differential privacy model than using all of the available features. For the UCI \breast{} dataset, this allows us to train a model with a $20.11\%$ relative increase in accuracy ($11.51\%$ absolute increase) than if we ignored the cost of achieving privacy. In some cases, using a small subset of {\em randomly} selected features yields a better model than using all available features. 

\paragraphb{Architecture search}
Taking the cost of privacy into account often results in simpler model architectures. On \cifar{}, we find a fully-connected architecture that has 32$\times$ fewer parameters than the more complex alternative, yet provides a 3.54\% absolute increase in accuracy when trained with \dpsgd{} for the same $\varepsilon$ (despite the more complex architecture having an $11.78\%$ absolute accuracy increase when trained with \sgd).  
In some cases, the difference is so large that different types of neural network architectures beat each other for the same task. For example, on \mnist{}, we discover a tiny fully-connected architecture with only $9,610$ parameters that outperforms the more complex convolutional architecture. More concerning perhaps, we find evidence that past wisdom in architecture design (without considering privacy) yields models that are inferior to simpler architectures designed with privacy in mind. For instance, we compare LeNet-5~\cite{lecun1998gradient} on \mnist{} with a CNN architecture we found. The latter provides (slightly) lower accuracy when the models are trained using \sgd{} ($99.28\%$ vs. $99.36\%$). However, when the models are trained with \dpsgd{} for $\varepsilon=1.09$, our architecture significantly outperforms LeNet-5 ($96.08\%$ vs. $94.30\%$).

\paragraphb{Towards best practices}
We demonstrate how the crossover epsilon can be used to choose a model architecture based on the available privacy budget. Our experiments also reveal unexpected actionable insights for practitioners. For example we find that the detrimental effect of DP noise is worse for deep neural networks than for wide neural networks (for the same number of trainable parameters). Another example is that an effective way to deal with stringent privacy guarantees (i.e., small $\varepsilon$) is to use a complex architecture with {\em random weights training}, i.e.: freeze the initial layers (after randomly initializing their weights) and train only the last few (e.g., 2) layers with \dpsgd. Bringing all of our experimental insights together, we derive a flowchart to help guide future ML engineering efforts.

\paragraphb{Privacy-aware algorithms}
We propose {\em privacy-aware} algorithms for feature and architecture selection. These algorithms are designed to take the effect of DP noise into account in a principled way. To account for cases where the search is based on sensitive data, we show that our algorithms satisfy differential privacy and analyze their privacy guarantees. We also evaluate empirically their privacy-utility tradeoffs and their running times.

\paragraphb{Structure}
We provide background on machine learning and differential privacy in~\cref{sec:background}. We describe our theory in~\cref{sec:theory}. We then turn to experiments and describe our setup and datasets in~\cref{sec:exp:setup}. We evaluate the pervasiveness and significance of the phenomenon in~\cref{sec:exp:demo} and then turn to validating our theory through experiments in~\cref{sec:exp:theoryval}. \cref{sec:exp:factorsflowchat} describes our efforts to uncover the factors that impact the phenomenon and distills our experimental insights into takeaways for practitioners in the form of a flowchart. In~\cref{sec:algos,sec:exp:algos} we describe our privacy-aware feature selection and architecture search algorithms and evaluate them experimentally. \cref{sec:related} provides a brief overview of related work and~\cref{sec:discussion} discusses the implications of our findings and outlines directions for future research efforts.

\section{Background \& Preliminaries}\label{sec:background}
In this section we provide background on supervised learning and differential privacy.

\subsection{Supervised Learning}
\label{sec:background:ml}
We consider a dataset $\ds$ of size $n>0$ where each data point $(\x_i, \y_i)$, $i \in [1, n]$ has been independently sampled from some (unknown) probability distribution $\dsdist$. Each $\x_i \in \inspace$ represents the features of the data point over the space $\inspace$ of features (e.g., $\inspace = \mathbb{R}^m$, where $m$ is the number of features) and each $\y_i \in \labspace$ represents either a class label (e.g., $\labspace = \{1, 2, \ldots, k\}$ where $k$ is the number of distinct classes) or a value (e.g., $\labspace = \mathbb{R}$). 

The model is defined by a function $\model : \inspace \to \outspace$ that is parameterized by a vector of real numbers $\params$. The output of the model for an input data point $\x \in \inspace$, denoted $\model(\x; \params)$, is the predicted class label or value $\lab \in \outspace$. We distinguish between the model architecture, denoted as $\model$, and the (trained) model alongside with its chosen parameter vector $\params$, which we denote as $\model_\params$ or $\model(\cdot; \params)$. 

Given the model architecture $\model$, training the model from dataset $\ds$ involves finding an optimal parameter vector $\params^\star$. For this we need to solve the following empirical risk minimization (ERM) problem: 
\begin{align} \label{eq:erm}
    \params^\star = \argmin_{\params \in \paramsspace} \sum_{i=1}^{n} \loss(\model(\x_i; \params), \y_i)  \ ,
\end{align}
where \paramsspace{} is the space of valid parameters and $l(\cdot)$ is a task-dependent loss function. There are multiple approaches to solve or approximate~\cref{eq:erm}, for example: Stochastic Gradient Descent (SGD)~\cite{gower2019sgd}.

\subsection{Differential Privacy}\label{sec:background:dp}
Differential privacy (DP)~\cite{dwork2006calibrating,dwork2014algorithmic} has become the de facto standard data privacy notion.
\begin{definition}
    A randomized algorithm \dpalg{} satisfies $(\varepsilon, \delta)$-{\em differential privacy} if for any neighboring datasets $\ds_0$, $\ds_1$ and any output set $S \subseteq {\rm Range}(\dpalg)$, we have:
        \[
            \pr{\dpalg(\ds_0) \in S} \leq \exp(\varepsilon) \ \pr{\dpalg(\ds_1) \in S} + \delta \ ,
        \]
\end{definition}
\noindent where the probabilities are taken over the randomness of \dpalg{}. Two datasets are neighboring if they differ in exactly one individual's record or data point. Here $\varepsilon > 0$ is called the {\em privacy budget} and is such that the smaller $\varepsilon$ is, the higher the privacy. Also, it is desirable to ensure that $\delta$ is (much) smaller than the reciprocal of the size of the input dataset (e.g., $\ll n^{-1}$). Whenever $\delta = 0$, we say that the algorithm satisfies {\em pure} differential privacy.

There are several {\em mechanisms} to transform an algorithm represented by a function $f$ into a randomized algorithm that satisfies $(\varepsilon, \delta)$-differential privacy. For example, the Laplace mechanism adds noise  from the Laplace distribution to the output of the algorithm. This noise must be calibrated to the (global) sensitivity of $f$, defined as: $\Delta_f = \max_{\ds_0, \ds_1} || f(\ds_0) - f(\ds_1) ||$, where the maximization is taken over all possible pairs of neighboring datasets $\ds_0, \ds_1$. 

For training machine learning models with differential privacy, there exists a drop-in replacement for SGD called Differentially Private Stochastic Gradient Descent (\dpsgd)~\cite{abadi2016deep}. Informally, \dpsgd{} computes the gradient at each iteration of \sgd{} and adds Gaussian noise to it. To ensure that this process satisfies DP despite the fact the sensitivity of the gradient is not easily bounded, the gradient is clipped prior to noise addition ensuring that its $L_2$-norm never exceeds a pre-defined threshold.

\subsection{Genetic Algorithms}\label{sec:background:genetic}
Genetic algorithms are a special case of evolutionary algorithms, which fit within the broader category of computational intelligence. These algorithms allow us to explore a large search space as a way to approximate an otherwise computationally infeasible optimization problem.

Informally, a genetic algorithm explores the space of solutions $Z$ to find the solution $z \in Z$ that meets or maximizes a specific criterion. The search proceeds in successive iterations called {\em generations}, wherein each generation the algorithm maintains a {\em population} of candidate solutions. We call a candidate solution an {\em individual}. A typical genetic algorithm has four components: (1) a fitness function that individuals are evaluated against; (2) a selection procedure that determines which individuals within the current population will survive to the next generation; (3) a crossover process that defines how two individuals may exchange ``genes'' to produce new candidate solutions; and (4) a mutation process that defines how an individual's genes mutate. When the algorithm terminates, the output is typically the candidate solution in the last generation's population that maximizes the fitness function.  

\section{Theoretical Framework}\label{sec:theory}

\subsection{An Explanatory Theory}
Suppose we are given a dataset $\ds$, a set of candidate model architectures $\modelspace$, and that we want to publish a model with the lowest possible generalization error but that is trained to satisfy ($\varepsilon,\delta$)-DP. We seek a simple theory that explains why the model architecture $h \in \modelspace$ that yields the lowest generalization error when trained without considering privacy is {\em not} necessarily the same one that yields the lowest generalization error when trained to satisfy ($\varepsilon,\delta$)-DP.

\paragraphb{Hypothesis} 
Our main hypothesis is that some model architectures that would yield low generalization errors without privacy are disproportionately hurt when trained to satisfy differential privacy. Furthermore: the more complex a model architecture (e.g., the larger its input and/or the more parameters it has), the more detrimental the effect of noise to achieve differential privacy. In other words, if we compare two model architectures $\model, \model' \in \modelspace$ where $\model'$ is derived from $\model$ with some architectural improvement (e.g., some additional features, an additional layer in the neural network, etc.) we expect that {\bf there is a tradeoff between the benefits of the architectural improvement in decreasing generalization error and its detrimental effect given differential privacy}.

More formally, given a model architecture $\model$, we care about its generalization error, which is the expected prediction error of $\model(\cdot; \params)$ over samples from the data distribution $\dsdist$, where $\params$ is obtained through a (possibly randomized) function $f$ that takes an input training dataset $\ds$, 
\begin{align*}
    \err{\model, f, \ds, \dsdist} = \mathbb{E}_{(\x, \y) \sim \dsdist, \ \params \sim f(\ds)}[\loss(\model(\x; \params), \y)] \ ,
\end{align*}
where the expectation is taken over both the data distribution $\dsdist$ and the randomness of $f$.

Without any privacy constraints, the function $f$ is an algorithm that approximately solves~\cref{eq:erm} such as \sgd. In contrast, if we seek to guarantee $(\varepsilon, \delta)$-differential privacy, the function $f$ is a randomized mechanism, denoted $F_{\varepsilon, \delta}$, such as \dpsgd.
For conciseness, we omit $\ds$, $\dsdist$, $\delta$, $f$, when they are clear from the context and simply write $\err{\model}$ for the generalization error with no privacy constraints and $\priverr{\varepsilon}{\model}$ for the generalization error of $\model$ under $(\varepsilon, \delta)$-differential privacy.

We define the {\em cost of achieving differential privacy} for a model architecture $\model$ and a privacy budget $\varepsilon$ as the difference between the generalization error when the parameters are obtained with $\varepsilon$-differential privacy and the generalization error with parameters obtained ignoring privacy. That is:
\begin{align}\label{eq:privcost}
    \cost{\varepsilon, \model} = \priverr{\varepsilon}{\model} - \err{\model} \ .
\end{align}
In practice $\cost{\varepsilon, \model}$ is usually positive, although it can sometimes be negative because differential privacy has a regularizing effect.\footnote{We also find in experiments (\cref{app:clip_gradients}) that even in case where the privacy budget is so large that the privacy guarantee is meaningless, training the model with \dpsgd{} sometimes yields a lower generalization error than training with \sgd{} because of gradient clipping.}

What is a meaningful measure of model architecture complexity? For the purpose of building intuition, we can think of complexity as the number of (trainable) parameters of the model. However, we show later that there are other factors at play. In the following subsection, we prove that our hypothesis holds for a specific DP mechanism (Gaussian noise output perturbation) and type of model (linear models).

\subsection{The Cost of Achieving Differential Privacy}\label{sec:theory:effectdp}
Consider a dataset with $m$ features and two potential alternative models: (1) a complete model that uses all $m$ features and (2) a simple(r) model that uses only the first $m-1$ features. If we train both models using traditional techniques and ignoring privacy, we expect the first model to provide better (or at least no worse) predictions than the second.\footnote{Even including irrelevant features will not harm accuracy in many cases. This is the case, for example, for logistic regression when using $L_1$ regularization~\cite{ng2004feature}.} However, we are concerned with what happens if we train both models with ($\varepsilon$,$\delta$)-differential privacy. 

To analyze this, we consider the Gaussian mechanism that first obtains the model's optimal parameter vector $\params$ through training and then adds isotropic Gaussian noise to it. In other words, the output of the mechanism is: 
\begin{equation}\label{eq:dpnoisegauss}
    \tilde{\params} = \params + {\bf z} \ ,
\end{equation}
where ${\bf z} \sim \mathcal{N}(0, \sigma^2 I)$ and $\sigma = \frac{\Delta}{\varepsilon} \sqrt{2 \ln{(1.25 / \delta)}}$. Here $\Delta$ denotes the $L_2$ sensitivity of the learning algorithm.

Remark that not all DP mechanisms add Gaussian noise this way. The output perturbation mechanism of Chaudhuri et al.~\cite{chaudhuri2011differentially} adds noise from a distribution with a different density. But as pointed by Wu et al.~\cite{wu2017bolt} noise from that distribution depends linearthmically on the dimension of the data which is undesirable, so they show how to use Gaussian noise instead. \dpsgd{} does add Gaussian noise, but the noise is added to the clipped gradient at each iteration and not directly to the learned parameter vector. In any case, our goal here is not to exhaustively analyze all the existing mechanisms, but rather to provide an illustrative derivation of the effect of DP noise.

What is the effect of Gaussian DP noise on the model's optimal parameter vector $\params$? Consider the distortion of $\tilde{\params}$ as in~\cref{eq:dpnoisegauss} compared to $\params$ due to the DP (Gaussian) noise. The $L_2$ norm of the noise follows a non-central Chi (aka generalized Rayleigh) distribution~\cite{miller1964distributions}, so the noise grows with $\sqrt{m}$. Similarly, the squared $L_2$ norm of the noise follows a Chi-square distribution with $m$ degrees of freedom~\cite{lancaster20020}, or equivalently a Gamma distribution with shape parameter $m/2$ and scale parameter $2 \sigma^2$. Thus, the expected squared $L_2$ norm of the noise is: $\mathbb{E}[{||{\bf z}||}_2^2] = \sigma^2 m$. This shows that the detrimental effect of (Gaussian) DP on the parameter vector grows with the number of features $m$. Further, this holds even if increasing the number of features does not (by itself) also increase the $L_2$ sensitivity $\Delta$ of the learning algorithm.

What about the effect of (Gaussian) DP noise on the model itself? We show how to analyze this effect when increasing the number of features, by making further assumptions. Consider two linear models $\model$ and $\model'$. The first model $\model$ has $m$ features and its optimal parameter vector is denoted by $\params$. The second model $\model'$ is identical to $\model$ but it excludes feature $m$ (so it uses only the first $m-1$ features) and we assume that its optimal parameter vector, denoted by $\params'$, is identical to $\params$ but with the coefficient corresponding to feature $m$ set to $0$. That is:
\begin{align*}
    \params = \begin{bmatrix}
           \theta_{1} \\
           \vdots \\
           \theta_{m-1} \\
           \theta_{m}
         \end{bmatrix}  
         \quad {\text{and}} \quad 
         \params' = \begin{bmatrix}
           \theta_{1} \\
           \vdots \\
           \theta_{m-1} \\
          0
         \end{bmatrix} . 
\end{align*}

We use the notation ${\bf x'}$ to denote feature vector ${\bf x}$ with its $m^{\rm th}$ row set to 0. We use the (squared) error of the model on $({\bf x},y)$ defined as: \[ {\rm err}(\params, {\bf x}, y) = (y - \params^T {\bf x})^2 = (y - \sum_i \theta_i x_i)^2 \ . \]

Consider the parameter vectors $\tilde{\params},\tilde{\params'}$ obtained after applying the Gaussian mechanism as in~\cref{eq:dpnoisegauss}. We have:
\begin{align*}
    \tilde{\params} = \begin{bmatrix}
           \theta_{1} + Z_1 \\
           \vdots \\
           \theta_{m-1} + Z_{m-1} \\
           \theta_{m} + Z_m
         \end{bmatrix}  
         \qquad 
    \tilde{\params'}  = \begin{bmatrix}
           \theta_{1} + Z_1'\\
           \vdots \\
           \theta_{m-1} + Z_{m-1}' \\
          0
         \end{bmatrix} \ ,
\end{align*}
where (by construction) $Z_i \sim \mathcal{N}(0, \sigma^2)$ for $i=1,2,\ldots,m$ and $Z_i' \sim \mathcal{N}(0, \sigma'^2)$. Note that there is no noise added to the $m^{\text{th}}$ coefficient of $\tilde{\params'}$ because the corresponding feature is excluded. Since the two models have a different number of features the $L_2$ sensitivity of the learning algorithm could be different (i.e., greater) for $\model$ than $\model'$ in which case we would have $\sigma' < \sigma$.
\begin{lemma}\label{lem:gaussiannoiselinearmodel}
    Let $\model,\model'$ be models as defined above. For any data point $({\bf x}, y)$ with ${\bf x} \neq 0$ and $x_m \neq 0$, then:
    \[
        \mathbb{E}[{\rm err}(\tilde{\params'}, {\bf x}, y)] \leq \mathbb{E}[{\rm err}(\tilde{\params}, {\bf x}, y)]
        \iff
        |\theta_m| \leq \frac{\sqrt{c^2 + d} - |c|}{|x_m|} \ ,
    \]
    where the expectation is taken over the randomness of the DP noise, $|c| = \sqrt{{\rm err}(\params, {\bf x}, y)}$, $d = a^2 - b^2$ for $a = \sigma ||{\bf x}||_2$ and  $b = \sigma' ||{\bf x'}||_2$. 
\end{lemma}
The proof of~\cref{lem:gaussiannoiselinearmodel} is deferred to~\cref{app:proofs}.

The following corollary provides a simpler bound on $|\theta_m|$ for the case where the non-DP model has $0$ error on a data point. It follows directly from~\cref{lem:gaussiannoiselinearmodel} by setting $c=0$ and observing that for $\sigma = \sigma'$ we have that $\sqrt{a^2 - b^2} = \sigma |x_m|$.
\begin{corollary}\label{cor:gaussiannoiselinearmodel}
    Let $\model,\model'$ be models as defined above and assume that $\sigma = \sigma'$. Then for any data point $({\bf x}, y)$ with ${\bf x} \neq 0$ (and $x_m \neq 0$) such that ${\rm err}(\params, {\bf x}, y) = 0$, we have:
    \[
        \mathbb{E}[{\rm err}(\tilde{\params'}, {\bf x}, y)] \leq \mathbb{E}[{\rm err}(\tilde{\params}, {\bf x}, y)]
        \iff  |\theta_m| \leq \sigma \ ,    
        \]
    where the expectation is taken over the randomness of the DP noise and $\sigma = \frac{\Delta}{\varepsilon} \sqrt{2 \ln{(1.25 / \delta)}}$.
\end{corollary}

\cref{cor:gaussiannoiselinearmodel} shows that adding feature $m$ only improves the model (under differential privacy) if the coefficient of the $m^{\text{th}}$ feature in the model trained without DP is greater in magnitude than the standard deviation of the DP noise. Although~\cref{lem:gaussiannoiselinearmodel} and~\cref{cor:gaussiannoiselinearmodel} apply to a specific setting (i.e., linear models with squared error and Gaussian DP noise), these results suggest that adding features is only beneficial when their incremental predictive power outweighs the detrimental effect of DP noise on the model.

\begin{figure}[!t]
\centering
    \includegraphics[width=0.995\linewidth]{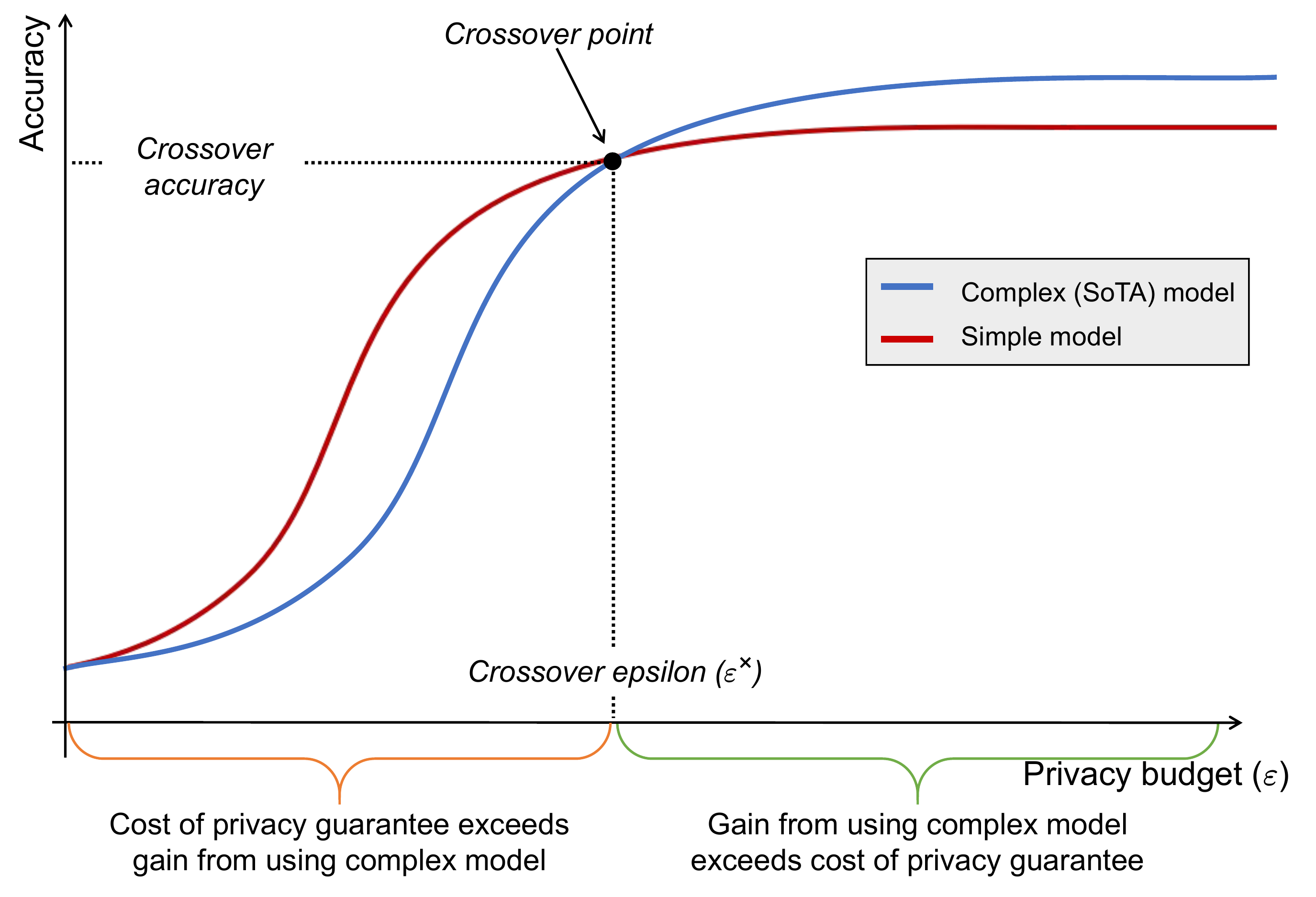}
    \caption{Illustration of the crossover epsilon.}
    \label{fig:illustration}
\end{figure}
\subsection{Deciding Between Models: Crossover Epsilon}\label{sec:framework:crossover}
How do we decide between several candidate model architectures? One way to answer this question is through a quantity that can be measured experimentally: the {\em crossover epsilon}. 

Suppose $\model_1$ and $\model_2$ are two model architectures for the same task and the same data distribution $\dsdist$. Given privacy budget ($\varepsilon$, $\delta$), we should prefer $\model_1$ whenever $\priverr{\varepsilon}{\model_1} < \priverr{\varepsilon}{\model_2}$. Otherwise we should use $\model_2$. 

From~\cref{eq:privcost}, we see that the only way for $\model_1$ to provide lower generalization error (than $\model_2$) given the privacy constraint is if:
\begin{align}\label{eq:invineq}
    \cost{\varepsilon, \model_1} < \cost{\varepsilon, \model_2} - [\err{\model_1} - \err{\model_2}] \ .
\end{align}
In other words, the cost of privacy for $\model_1$ must be smaller than the difference between the cost of privacy for $\model_2$ and the generalization error gap of the two models when ignoring privacy.

For our purposes, we assume that given two model architectures, we always denote the simple(r) model as $\model_1$ and the (more) complex model as $\model_2$. For example, $\model_2$ may be the state-of-the-art architecture for a given task and $\model_1$ may be a much simpler alternative. Given this, we expect that $\err{\model_2} \leq \err{\model_1}$, that is the complex model performs better than the simple model when ignoring privacy. Assuming that this is the case~\cref{eq:invineq} can only hold if the privacy cost of $\model_1$ is smaller than that of $\model_2$, i.e.: $\cost{\varepsilon, \model_1} < \cost{\varepsilon, \model_2}$. This is a necessary but not sufficient condition. 

Now assume that the cost of privacy $\cost{\varepsilon, \model}$ for any $\model$ decreases as $\varepsilon$ increases. Also, as $\varepsilon \to 0$ the generalization error $\priverr{\varepsilon}{\model}$ approaches the error of random guessing.  If there exists a (possibly large) $\varepsilon' > 0$ such that $\priverr{\varepsilon}{\model_2} > \priverr{\varepsilon}{\model_1}$ for any $\varepsilon > \varepsilon'$, then there must also exist  $\varepsilon'' < \varepsilon'$ such that $\priverr{\varepsilon''}{\model_1} = \priverr{\varepsilon''}{\model_2}$. In plain words: the curves of generalization error of the two models --- viewed as functions of $\varepsilon$ given the privacy constraint --- must cross. We call the crossing point the {\em crossover epsilon}.
\begin{definition}[Crossover $\varepsilon$]
    Given $\model_1$ and $\model_2$ such that $\err{\model_1} > \err{\model_2}$, the {\em crossover epsilon} is: 
    \begin{align}
        \varepsilon^\times = \varepsilon^\times(\model_1, \model_2) = \begin{cases}
                                \max(S)  & \text{if } S \neq \emptyset \\
                                \infty & \text{otherwise}
                            \end{cases}
                            \,
    \end{align}
    where $S = \{ \varepsilon : \priverr{\varepsilon}{\model_1} < \priverr{\varepsilon}{\model_2} \ \text{ for } 0 < \varepsilon < \infty \}$.
\end{definition}

The crossover epsilon is significant because it helps us determine which model architecture to use depending on our privacy budget $\varepsilon$: whenever $\varepsilon \leq \varepsilon^\times$ we should use $\model_1$ otherwise we should use $\model_2$. This is illustrated in~\cref{fig:illustration}. Note that since we do not make any assumptions on the mechanism used to achieve differential privacy, it is possible that the two curves $\priverr{\varepsilon}{\model_1}$ and $\priverr{\varepsilon}{\model_2}$ (viewed as functions of $\varepsilon$) cross each other multiple times. However, we have not observed this behavior in experiments. 

While the crossover epsilon can help us decide between two model architectures given our privacy budget, a more principled way to approach the problem is to discover architectures in a privacy-aware way.

\subsection{Privacy-Aware Search and Workflows}\label{sec:framework:problem}
We propose a reformulation of the optimization problem of~\cref{eq:erm} to take into account feature selection and architecture search. Let $\modelspace$ denote the space of all considered model architectures. In this paper, we use model architecture to refer to model types, selected features, hyperparameters, and neural network architecture (if the model is a neural network).
\begin{align} \label{eq:optprob}
    (\model^\star, \params^\star) = \argmin_{\model \in \modelspace, \ \params \in \paramsspace(\model)} \ \sum_{i=1}^{n} \loss(\model_\params(\x_i), \y_i)  \ , 
\end{align}
where the output is now a pair consisting of model architecture $\model^\star$ and its corresponding optimal parameter vector $\params^\star$. Note that the optimal parameter vector $\params^\star$ can only be determined given a specific model architecture.

The framing of~\cref{eq:optprob} allows us to capture both feature selection and model architecture selection. In particular, since $\model \in \modelspace$ may be such that it ignores certain features or combines features together, it can encapsulate some feature engineering operations. Further, a specific $\model \in \modelspace$ can be thought of as encapsulating its own hyperparameter values and architecture.

\paragraphb{The standard (or ``drop-in'') workflow (STW)}
Given a (non private) procedure to solve~\cref{eq:erm} given a dataset $\ds$, a simple approach for releasing a differential private machine learning model is to first use the procedure to obtain a solution $(\model^\star, \params^\star)$ of~\cref{eq:optprob} and then use a differential privacy mechanism $\mc{F}_{\varepsilon,\delta}$ to train $\model^\star$ on dataset $\ds$ to obtain parameters $\tilde{\params}$. Then, one simply releases the tuple $(\model^\star, \tilde{\params})$. We call this approach the {\em standard (or ``drop-in'') workflow} (STW). It maps onto the workflow of a practitioner who performs all operations without considering privacy until a suitable architecture is identified and then uses a differentially private learning algorithm to train the model. Arguably this is also the workflow often used in research that evaluates various aspects of differential privacy machine learning algorithms~\cite{abadi2016deep,bagdasaryan2019differential,song2013stochastic} or the use of differential privacy as a defense for inference attacks~\cite{rahman2018membership,he2020segmentations,zhang2022label}.

\paragraphb{The privacy-aware workflow (PAW)}
By contrast to the standard workflow, a principled way to solve~\cref{eq:optprob} is to take into account the cost of achieving differential privacy in the selection of model type, architectures, and features. We call this the {\em privacy-aware workflow} (PAW). In the last part of this paper, we propose privacy-aware algorithms for feature selection and architecture search. We defer their description to~\cref{sec:algos} and focus for the next few sections on explicating and demonstrating the pitfall of using the standard workflow.

\paragraphb{Differential privacy of the workflow}
Although the released model when we follow the standard workflow or the privacy-aware workflow will satisfy $(\varepsilon,\delta)$-differential privacy, the workflow as a whole may not satisfy differential privacy. This is because choices that are made as part of the workflow such as the selection of the model architecture could be based on sensitive data without accounting for that leakage. We set this issue aside for now and defer the discussion of how to ensure that our privacy-aware algorithms satisfy differential privacy in~\cref{sec:algos:dpsearch}.

\section{Datasets and Experiments setup}\label{sec:exp:setup}
\subsection{Datasets}\label{app:data}
In this section, we describe the datasets we use and how we pre-process them.

\paragraphb{\cifar}
\cifar{} is a computer vision data set for universal object recognition that is commonly used as a benchmark for image recognition classes. It was collected by Alex Krizhevsky, Vinod Nair, and Geoffrey Hinton~\cite{krizhevsky2009learning}. The dataset contains 60,000 color images classified into 10 classes (cat, dog, airplane, and so on). Each image is $32 \times 32$. We use 50,000 data points for training, 5,000 for validation, and 5,000 for the test set.

\paragraphb{\mnist}
\mnist{}~\cite{lecun1998mnist} is a dataset of handwritten digit (0 through 9) images. The dataset contains 70,000 28 $\times$ 28 gray-scale images. Identifying the number shown in the image is a classic image classification task. We use 60,000 data points for training, 5,000 for validation, and 5,000 for the test set.

\paragraphb{\fmnist}
\fmnist{} is a dataset of 28 $\times$ 28 gray-scale images of clothing items (e.g. shirts, dresses, bags, shoes, etc.) provided by the research division of Zalando (a German fashion technology company)~\cite{xiao2017fashion}. The dataset contains 70,000 images, divided into sets of 60,000, 5,000, and 5,000 for training, validation, and test sets respectively.

\paragraphb{\svhn}
The Street View House Numbers (\svhn) dataset was extracted from Google Street View images of door signs~\cite{netzer2011reading}. It contains 52,000 images (32$\times$32). In this paper, we use a training set of 42,000 images and validation and test sets of 5,000 data images each. Also, we only use gray-scale images.

\paragraphb{\adult}
This is a dataset hosted on the UCI Machine Learning repository.\footnote{\url{https://archive.ics.uci.edu/ml}} It is often used to predict whether a person's income is over 50K USD or not~\cite{kohavi1996scaling} on the basis of 14  (mostly categorical) attributes. The data was extracted from the 1994 census. We preprocessing the data by one-hot encoding all attributes. This results in 108 features. We use 22,750 records in the training data and 9,750 for the test set. 

\paragraphb{\purchase}
This dataset is based on Kaggle's ``acquire valued shoppers'' challenge.\footnote{\url{https://www.kaggle.com/c/acquire-valued-shoppers-challenge}} The dataset was prepared through clustering by Shokri et al.~\cite{shokri2017membership} to have 100 classes. Each dataset record has 600 binary features and there are 197,000 records in total. We use 157,750 as the training set, 5,000 as the validation set, and 5,000 as the test set.

\paragraphb{\breast}
This is a dataset hosted on the UCI Machine Learning repository \cite{breast_cancer}. It is often used to predict whether there are recurrence events. It has 9 attributes and most of them are categorical. We use one-hot encoding to prepossess all attributes and results in 43 attributes. We use 191 records to train and 95 records to test.

\paragraphb{\spam}
This is a dataset hosted on the UCI Machine Learning repository \cite{sms_spam}. 
It is often used to predict whether this message is spam or not. It contains 5,574 records (English message and their labels). We use 4700 records to train the RNN model and 836 records for testing (we ignore some data records to make sure the training dataset can be evenly divided by batch size which is required by \dpsgd{}).

\paragraphb{\medical} 
This is a dataset pulled from Machine Learning with R by Brett Lantz~\cite{lantz_2015} and hosted on Kaggle~\cite{choi_2018}. This dataset contains basic attributes of patients, such as age, BMI, and number of children, and is used to predict the individual medical cost charged by the patient's insurance. It contains 1338 entries, 1000 of which we use to train regression models, and 300 of which we use for testing.

\subsection{Setup}\label{app:setup}
We used NVIDIA DGX A100 GPUs for architecture search and training models. We used Tensorflow 2.4.0 with Python 3.8 and the latest version at the time of writing (i.e., version 0.6.2) of the tensorflow-privacy package for \dpsgd{}.\footnote{\url{https://github.com/tensorflow/privacy}} To do the accounting of privacy parameters ($\varepsilon$, $\delta$), we use the \textsf{compute\_dp\_sgd\_privacy()} function. We use grid search to find the most suitable learning rate and other hyperparameters for training neural networks.

Unless otherwise stated, the differential privacy settings are as follows. The noise level for \dpsgd{} is 2.0, L2 norm is set to 1.0, and $\delta \leq 10^{-5}$ for all datasets expect for \purchase{} (where $\delta \leq 10^{-7}$ due to the larger number of records in this dataset).

\begin{table*}[ht!]
\caption{Summary results for FCNs and CNNs for STW and PAW models (we set $\delta=10^{-7}$ for \purchase{} and $\delta=10^{-5}$ for all other datasets). Gap is defined as the difference in test accuracy between the PAW model and the STW model. PR denotes the ratio of STW model parameters to PAW model parameters. In all cases the PAW model outperforms the STW model when training with \dpsgd. In contrast, the STW model provides significantly higher accuracy when trained with \sgd{} (i.e., with no privacy).
\label{tbl:arch-gensearch}}
\vspace{-2pt}
\centering
\resizebox{0.805\linewidth}{!}{%
\begin{tabular}{cccc|c|c|c|c|cc}
\cline{5-8}
 &
   &
   &
   &
  \multicolumn{2}{c|}{Test Accuracy} &
  \multicolumn{2}{c|}{Gap} &
   &
   \\ \hline
\multicolumn{1}{|c|}{Architecture} &
  \multicolumn{1}{c|}{Dataset} &
  \multicolumn{1}{c|}{Workflow} &
  Parameters &
  DP-SGD &
  SGD &
  DP-SGD &
  SGD &
  \multicolumn{1}{c|}{PR} &
  \multicolumn{1}{c|}{$\varepsilon$} \\ \hline \hline
\multicolumn{1}{|c|}{\multirow{6}{*}{CNN}} &
  \multicolumn{1}{c|}{\multirow{2}{*}{\cifar}} &
  \multicolumn{1}{c|}{STW} &
  1,820,330 &
  47.76\% &
  \textbf{67.54\%} &
  \multirow{2}{*}{2.46\%} &
  \multirow{2}{*}{-6.08\%} &
  \multicolumn{1}{c|}{\multirow{2}{*}{7.4$\times$}} &
  \multicolumn{1}{c|}{\multirow{2}{*}{4.87}} \\ \cline{3-6}
\multicolumn{1}{|c|}{} &
  \multicolumn{1}{c|}{} &
  \multicolumn{1}{c|}{PAW} &
  246,554 &
  \textbf{50.22\%} &
  61.46\% &
   &
   &
  \multicolumn{1}{c|}{} &
  \multicolumn{1}{c|}{} \\ \cline{2-10} 
\multicolumn{1}{|c|}{} &
  \multicolumn{1}{c|}{\multirow{2}{*}{\mnist}} &
  \multicolumn{1}{c|}{STW} &
  197,130 &
  96.88\% &
  \textbf{99.42\%} &
  \multirow{2}{*}{0.98\%} &
  \multirow{2}{*}{-0.04\%} &
  \multicolumn{1}{c|}{\multirow{2}{*}{3.5$\times$}} &
  \multicolumn{1}{c|}{\multirow{2}{*}{4.39}} \\ \cline{3-6}
\multicolumn{1}{|c|}{} &
  \multicolumn{1}{c|}{} &
  \multicolumn{1}{c|}{PAW} &
  56,970 &
  \textbf{97.86\%} &
  99.38\% &
   &
   &
  \multicolumn{1}{c|}{} &
  \multicolumn{1}{c|}{} \\ \cline{2-10} 
\multicolumn{1}{|c|}{} &
  \multicolumn{1}{c|}{\multirow{2}{*}{\svhn}} &
  \multicolumn{1}{c|}{STW} &
  1,817,930 &
  82.72\% &
  \textbf{91.68\%} &
  \multirow{2}{*}{2.00\%} &
  \multirow{2}{*}{-0.78\%} &
  \multicolumn{1}{c|}{\multirow{2}{*}{3.0$\times$}} &
  \multicolumn{1}{c|}{\multirow{2}{*}{5.38}} \\ \cline{3-6}
\multicolumn{1}{|c|}{} &
  \multicolumn{1}{c|}{} &
  \multicolumn{1}{c|}{PAW} &
  600,522 &
  \textbf{84.72\%} &
  90.90\% &
   &
   &
  \multicolumn{1}{c|}{} &
  \multicolumn{1}{c|}{} \\ \hline \hline
\multicolumn{1}{|c|}{\multirow{10}{*}{FCN}} &
  \multicolumn{1}{c|}{\multirow{2}{*}{\cifar}} &
  \multicolumn{1}{c|}{STW} &
  636,938 &
  40.12\% &
  \textbf{57.14\%} &
  \multirow{2}{*}{3.54\%} &
  \multirow{2}{*}{-11.78\%} &
  \multicolumn{1}{c|}{\multirow{2}{*}{32.0$\times$}} &
  \multicolumn{1}{c|}{\multirow{2}{*}{2.34}} \\ \cline{3-6}
\multicolumn{1}{|c|}{} &
  \multicolumn{1}{c|}{} &
  \multicolumn{1}{c|}{PAW} &
  19,914 &
  \textbf{43.66\%} &
  45.36\% &
   &
   &
  \multicolumn{1}{c|}{} &
  \multicolumn{1}{c|}{} \\ \cline{2-10} 
\multicolumn{1}{|c|}{} &
  \multicolumn{1}{c|}{\multirow{2}{*}{\mnist}} &
  \multicolumn{1}{c|}{STW} &
  66,762 &
  93.10\% &
  \textbf{97.54\%} &
  \multirow{2}{*}{1.84\%} &
  \multirow{2}{*}{-1.20\%} &
  \multicolumn{1}{c|}{\multirow{2}{*}{6.9$\times$}} &
  \multicolumn{1}{c|}{\multirow{2}{*}{2.11}} \\ \cline{3-6}
\multicolumn{1}{|c|}{} &
  \multicolumn{1}{c|}{} &
  \multicolumn{1}{c|}{PAW} &
  9,610 &
  \textbf{94.94\%} &
  96.34\% &
   &
   &
  \multicolumn{1}{c|}{} &
  \multicolumn{1}{c|}{} \\ \cline{2-10} 
\multicolumn{1}{|c|}{} &
  \multicolumn{1}{c|}{\multirow{2}{*}{\svhn}} &
  \multicolumn{1}{c|}{STW} &
  221,066 &
  72.90\% &
  \textbf{85.32\%} &
  \multirow{2}{*}{3.20\%} &
  \multirow{2}{*}{-1.20\%} &
  \multicolumn{1}{c|}{\multirow{2}{*}{5.6$\times$}} &
  \multicolumn{1}{c|}{\multirow{2}{*}{2.58}} \\ \cline{3-6}
\multicolumn{1}{|c|}{} &
  \multicolumn{1}{c|}{} &
  \multicolumn{1}{c|}{PAW} &
  39,818 &
  \textbf{76.10\%} &
  84.12\% &
   &
   &
  \multicolumn{1}{c|}{} &
  \multicolumn{1}{c|}{} \\ \cline{2-10} 
\multicolumn{1}{|c|}{} &
  \multicolumn{1}{c|}{\multirow{2}{*}{\fmnist}} &
  \multicolumn{1}{c|}{STW} &
  38,410 &
  85.00\% &
  \textbf{89.62\%} &
  \multirow{2}{*}{1.74\%} &
  \multirow{2}{*}{-1.92\%} &
  \multicolumn{1}{c|}{\multirow{2}{*}{4.0$\times$}} &
  \multicolumn{1}{c|}{\multirow{2}{*}{2.11}} \\ \cline{3-6}
\multicolumn{1}{|c|}{} &
  \multicolumn{1}{c|}{} &
  \multicolumn{1}{c|}{PAW} &
  9,610 &
  \textbf{86.74\%} &
  87.70\% &
   &
   &
  \multicolumn{1}{c|}{} &
  \multicolumn{1}{c|}{} \\ \cline{2-10} 
\multicolumn{1}{|c|}{} &
  \multicolumn{1}{c|}{\multirow{2}{*}{\purchase}} &
  \multicolumn{1}{c|}{STW} &
  903,524 &
  66.74\% &
  \textbf{94.88\%} &
  \multirow{2}{*}{8.04\%} &
  \multirow{2}{*}{-3\%} &
  \multicolumn{1}{c|}{\multirow{2}{*}{10.1$\times$}} &
  \multicolumn{1}{c|}{\multirow{2}{*}{1.54}} \\ \cline{3-6}
\multicolumn{1}{|c|}{} &
  \multicolumn{1}{c|}{} &
  \multicolumn{1}{c|}{PAW} &
  89,828 &
  \textbf{74.78\%} &
  91.88\% &
   &
   &
  \multicolumn{1}{c|}{} &
  \multicolumn{1}{c|}{} \\ \hline
\end{tabular}%

}
\vspace{-8pt}
\end{table*}
\section{Evaluation: Pervasiveness \& Impact}\label{sec:exp:demo}
We seek to evaluate how widespread the phenomenon is and quantify its impact. 

\subsection{Architecture Search}\label{sec:eval:arch}

\paragraphb{Setup \& Methodology} \label{sec:eval:arch:setup}
For each dataset/prediction task, we identify suitable model architectures for the standard workflow and the privacy-aware workflow. (The details of the search algorithms we use are deferred to~\cref{sec:algos}.) We then train the models and compare their test accuracy in the non-private and differential privacy cases.
We use fully-connected feed-forward neural networks (FCNs) and convolutional neural networks (CNNs) that we train with \sgd{} and \dpsgd{}. When comparing STW and PAW models, we use \dpsgd{} ensuring that the privacy parameters ($\varepsilon,\delta$) are the same in both cases to obtain a fair comparison. We also identify a close to optimal learning rate in each case. For both FCN and CNN experiments, we repeat each experiment at least three times to overcome variance from random fluctuations. We report the best results from each experiment based on the test accuracy.

\paragraphb{Results} \label{sec:eval:arch:result}
We perform experiments on five datasets using two types of neural networks. The results are summarized in~\cref{tbl:arch-gensearch} (details of architectures used can be found~\cref{app:arch}). We see that the generalization error ---measured as test accuracy on $\varepsilon$-differential privacy model--- is significantly lower for the privacy-aware workflow (PAW) model than that of the standard workflow (STW) model. At the same time, when ignoring privacy (training with \sgd) the STW model's performance is significantly better than that of the PAW model. This is exactly what we expect when~\cref{eq:invineq} holds.

To emphasize the point: consider the STW CNN architecture on CIFAR-10. Achieving differential privacy results in an absolute decrease in accuracy of almost $20\%$. By contrast, achieving differential privacy for the PAW architecture incurs an absolute decrease of only $11\%$. The consequence of this is that {\bf despite the STW architecture being superior for CIFAR-10 than the PAW architecture, the latter outperforms the former given differential privacy.}

In all the cases in~\cref{tbl:arch-gensearch}, we find large generalization error gaps. The largest absolute gap is observed for the \purchase{} dataset ($8.04\%$). Furthermore, we observe that the PAW architectures have significantly fewer parameters than the corresponding STW architectures. In the most striking case, the FCN PAW architecture for CIFAR-10 has $32 \times$ fewer parameters. That said, in our experiments we have not observed the reduction in parameters to be (by itself) predictive of the generalization error gap.

\paragraphq{Is this limited to classification and/or FCN and CNN architectures?} \label{sec:eval:arch:othertype}
No. We find that the same phenomenon occurs for all of the tasks and all architecture types that we have considered. In particular, the phenomenon manifests the same way for regression tasks and when using recurrent neural networks. 
We provide additional experiments in~\cref{app:exp}. Taken together, our results strongly suggest that the phenomenon is commonplace and is not limited to any types of model architectures or tasks.

\begin{figure*}[t!]
\centering
\begin{minipage}[b]{.475\textwidth}
\centering
\includegraphics[width=0.885\linewidth]{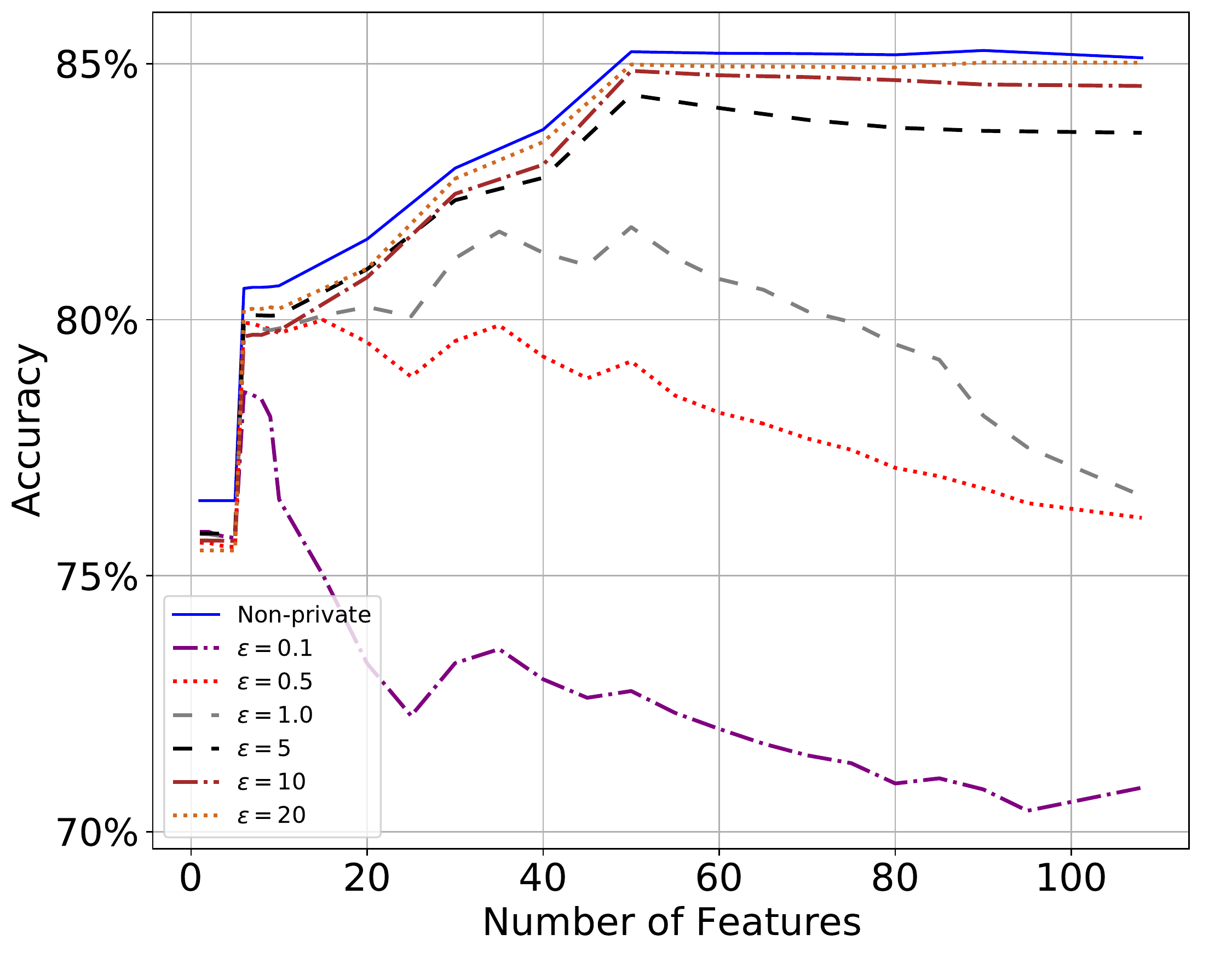}
\vspace{-3pt}
\caption{Logistic regression models for the \adult{} dataset trained with $\varepsilon$-differential privacy (objective perturbation) over the best subset of features of each size as identified by our privacy-aware greedy feature selection algorithm. With no privacy constraints, the best model uses all features. For stringent privacy constraints ($\varepsilon = 0.1$) the best model is one that uses only 5 features. }
\vspace{-8pt}
\label{fig:cfs_feature_acc_adult}
\end{minipage}\qquad
\begin{minipage}[b]{.475\textwidth}
\centering
\includegraphics[width=0.885\linewidth]{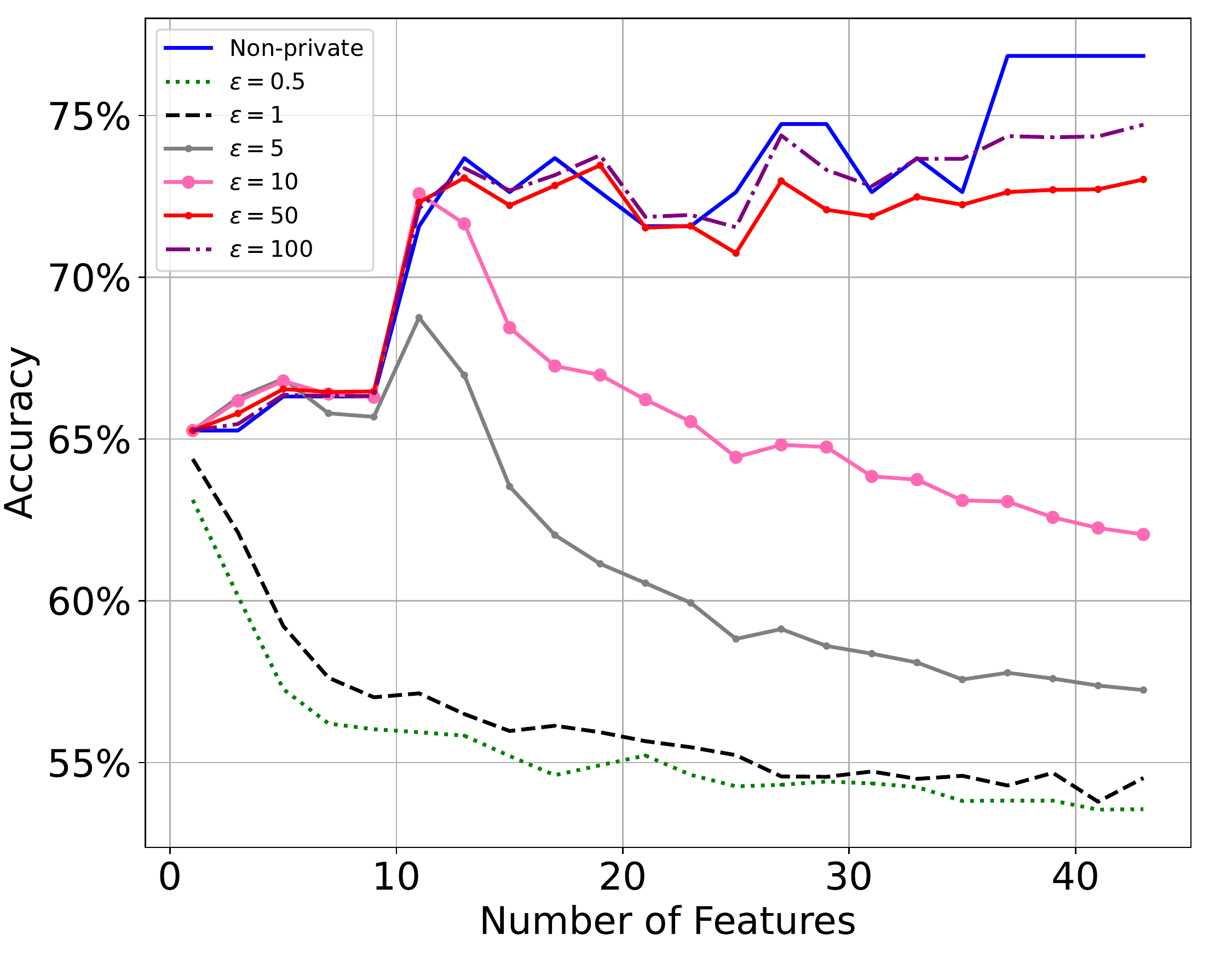}
\vspace{-3pt}
\caption{Logistic regression models for the \breast{} dataset trained with $\varepsilon$-differential privacy (objective perturbation) over the best subset of features of each size as identified by our privacy-aware greedy feature selection algorithm. With no privacy constraints, the best model is the one that uses all features. For $\varepsilon = 5$ the best model is one that uses only 11 features.}
\vspace{-8pt}
\label{fig:cfs_feature_acc_breast}
\end{minipage}
\end{figure*}
\subsection{Feature Selection}
We now turn to the case of feature selection. We are interested in whether the use of a subset of features yields a better model with differential privacy than using all available features. 

\paragraphb{Setup \& Methodology}\label{sec:eval:fea:setup}
We focus on logistic regression classifiers with the \adult{} and \breast{} datasets. We use the objective perturbation technique proposed by Chaudhuri et al.~\cite{chaudhuri2011differentially} as implemented in Diffprivlib (the IBM differential privacy library).\footnote{\url{https://github.com/IBM/differential-privacy-library}} To obtain a baseline where there is no privacy constraint, we also train the models with a {\em very large} privacy budget, i.e.: $\varepsilon = 10^{9}$. In both cases, we set the maximum number of iterations for the solver to 5,000 and the tolerance to $10^{-50}$. We also use \sgd{} and \dpsgd{} with the setup described in~\cref{sec:eval:arch:setup} for some experiments. 

To reduce the impact of random fluctuations in our results, we repeat experiments with objective perturbation at least 1000 times and report the average accuracy on the test dataset. For \dpsgd{} and \sgd{}, we also repeat the \dpsgd{} and \sgd{} training at least 10 times and report the average test accuracy.

\paragraphq{Is using more features always better?}\label{sec:eval:fea:more_features}
We use a greedy approach to identify the best subsets of features of each size from $1$ to $108$ for \adult{} and $1$ to $43$ for \breast{}. (The details of the algorithms are presented in~\cref{sec:algorithms:fs}.) We then train logistic regression models for all of these subsets while varying the privacy budget $\varepsilon$ from $0.1$ to $20$ for \adult{} and $0.5$ to $100$ for \breast{} with objective perturbation. 

The results are shown in~\cref{fig:cfs_feature_acc_adult,fig:cfs_feature_acc_breast}. We see that for stringent privacy constraints (i.e., small privacy budget $\varepsilon$) the model's accuracy decreases the more features are used. In contrast when the model is trained with no privacy constraints (blue lines), the more features are used the higher the accuracy (except in some cases for \breast{}). This point is made more salient by noticing the large accuracy decrease of using all features as opposed to a small carefully selected set of features when the privacy budget is small (e.g., $\varepsilon \leq 0.5$ for \adult{} and $\varepsilon \leq 1.0$ for \breast{}). For example, for \adult{} with $\varepsilon = 0.1$ using all features gives us a model with $70.86\%$ accuracy, whereas using the best subset of five features yields a model with $78.58\%$ accuracy. In addition this effect is particularly pronounced for the \breast{} models where for $\varepsilon \leq 1$ the best model uses the single best feature.

Furthermore, for moderate values of the privacy budget (e.g., $\varepsilon = 1.0$ for \adult{} and $\varepsilon = 5$ for \breast{}) the best performing model is obtained with a strict subset of features in the middle of the range. Indeed, in such cases {\bf using all available features results in a worse performing model than only using a small carefully selected subset}. This is consistent with our theoretical results (\cref{sec:theory:effectdp}).

\pagebreak
\paragraphq{Does the DP mechanism matter?}\label{sec:eval:arch:mechanism}
Is the behavior dependent on which $\varepsilon$-differential privacy mechanism is used? To answer this question, we repeat the feature selection experiments using \dpsgd{} instead of objective perturbation~\cite{chaudhuri2011differentially}. In this case we use a neural network with a single layer of two neurons (essentially implementing binary logistic regression as a neural network). We then train it with \sgd{} and \dpsgd{} for two values of $\varepsilon$ over the \adult{} and \breast{} datasets while varying the number of features used. In this case, we consider the best subsets of features of five different sizes and comparing this to using all available features. We train models for $40$ epochs and with a batch size of $250$. We use categorical cross-entropy as the loss function. Results are shown in~\cref{tbl:cfs_sgd_adult} (\adult{}) and~\cref{tbl:cfs_sgd_breast} (\breast{}). We see that for the chosen privacy budgets $\varepsilon$, using a subset of features provides a more accurate model than using all features. However, in some cases (e.g., \adult{} with 2 features for $\varepsilon=0.0196$) using too few features provides a worse performing model. The privacy budget $\varepsilon$ in this case is much smaller than for objective perturbation, because \dpsgd{} provides superior privacy guarantees for the same accuracy.

\vspace{-4pt}
\subsection{Takeaways}\label{sec:eval:takeaway1}
Following the standard ``drop-in'' workflow ignores the cost of achieving differential privacy during architecture search and feature selection. We find that this results in systematically using overly complex models that make poor predictions, where there exist significantly better performing alternatives. These alternatives, which are readily found by accounting for differential privacy in the search, yield consistently better predictions and use fewer parameters (sometimes by an order of magnitude). This behavior is observed consistently across all types of model architectures (FCN, CNN, RNNs), types of tasks (classification and regression), and all datasets we have tried. Furthermore, it is not specific to the mechanism used to achieve differential privacy.
\begin{table}[!t]
\centering
\caption{Test accuracy for feature selection with \dpsgd{} and \sgd{} for \adult{}. With \sgd{} using all features always results in higher test accuracy, whereas with \dpsgd{} using a small subset of features provides a better model, especially for small $\varepsilon$.}
\label{tbl:cfs_sgd_adult}
\resizebox{\linewidth}{!}{%
\begin{tabular}{cc|c|c|c|c|}
\cline{3-6}
                                     &                & \multicolumn{2}{c|}{Selected features} & \multicolumn{2}{c|}{All features} \\ \hline
\multicolumn{1}{|c|}{$\varepsilon$} & \# of Features & SGD                & DP-SGD            & SGD            & DP-SGD           \\ \hline
\multicolumn{1}{|c|}{\multirow{5}{*}{0.236}} & 2 & 65.26\% & 67.37\% & \multirow{10}{*}{73.68\%} & \multirow{5}{*}{66.32\%} \\ \cline{2-4}
\multicolumn{1}{|c|}{}               & 10             & 69.47\%            & 65.26\%           &                &                  \\ \cline{2-4}
\multicolumn{1}{|c|}{}               & 20             & 72.63\%            & 68.42\%           &                &                  \\ \cline{2-4}
\multicolumn{1}{|c|}{}               & 30             & 71.58\%            & 66.31\%           &                &                  \\ \cline{2-4}
\multicolumn{1}{|c|}{}               & 40             & 73.68\%            & 66.26\%           &                &                  \\ \cline{1-4} \cline{6-6} 
\multicolumn{1}{|c|}{\multirow{5}{*}{0.014}} & 2 & 65.26\% & 65.26\% &                           & \multirow{5}{*}{59.48\%} \\ \cline{2-4}
\multicolumn{1}{|c|}{}               & 10             & 69.47\%            & 64.21\%           &                &                  \\ \cline{2-4}
\multicolumn{1}{|c|}{}               & 20             & 72.63\%            & 60.00\%           &                &                  \\ \cline{2-4}
\multicolumn{1}{|c|}{}               & 30             & 71.58\%            & 62.11\%           &                &                  \\ \cline{2-4}
\multicolumn{1}{|c|}{}               & 40             & 73.68\%            & 59.35\%           &                &                  \\ \hline
\end{tabular}
}
\vspace{-8pt}
\end{table}

\section{Evaluation: Theory Validation}\label{sec:exp:theoryval}
We perform experiments to validate our theoretical framework (\cref{sec:theory}).

\subsection{Analyzing the effect of DP noise}\label{sec:eval:dp-noise}
We devise an experiment to analyze the effect of DP noise on models of varying complexity. For this we engineer a dataset and prediction task and compare models with varying input sizes. Specifically, we generate random feature values in $[0,1]$ and label each data point as the sum of the feature values. To vary the input size (i.e., number of features) we simply divide the original feature value to create multiple new features which add up to the original value. Using this method we can guarantee that the relationship between input and label is (essentially) the same, which ensures that models trained with different input sizes will effectively learn the same relationship and achieve the same accuracy. We use single-layer neural networks with a single output unit using sigmoid activation but five different input size/complexity (i.e. 1000, 500, 100, 50, and 10). We set the logistic loss which is strictly convex and thus ensures that there is a unique optimal set of parameters in each case.

\begin{figure}[!t]
\centering
    \includegraphics[width=0.95\linewidth]{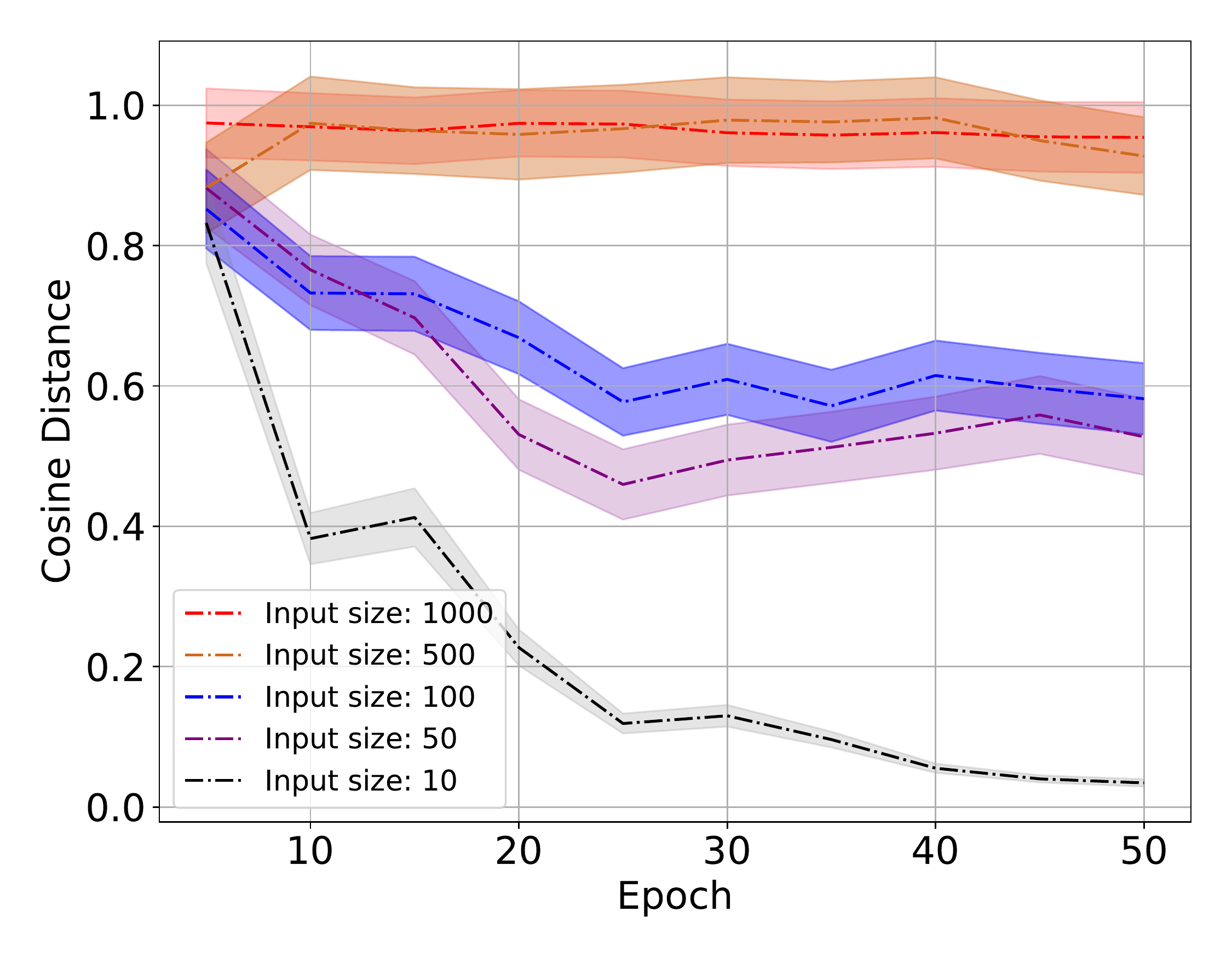}
    \caption{Cosine distance of model parameters with SGD and DP-SGD for different input sizes. The shaded area around each curve shows a 95\%-confidence interval.}
    \label{fig:explanation_ci}
    \vspace{-6pt}
\end{figure}

We train the models using both \sgd{} and \dpsgd{} to understand the effect of the noise. Specifically, for each model we analyze what happens to the parameters during training by measuring the cosine distance of model parameters with \sgd{} and \dpsgd{}. Because of the way the task is engineered: all models will achieve 100\% accuracy when trained with \sgd{}, because there is only one optimal set of parameters in each case and it is the same for \sgd{} and \dpsgd{}. To overcome the randomness of the initialization and the training process, we repeat this experiment 100 times. \cref{tab:explanation_test_accuracy} shows the test accuracy for \sgd{} and \dpsgd{} of the models for each input size. \cref{fig:explanation_ci} shows the cosine distance measured across 50 training epochs. The shaded region around each line shows an approximate 95\%-confidence interval. 
\begin{table}[t!]
\centering
\caption{Detrimental effect of differential privacy noise. Test accuracy for different model input sizes for the same task.}
\vspace{-3pt}
\label{tab:explanation_test_accuracy}
\begin{tabular}{cc|cc|}
\cline{3-4}
                          &  & \multicolumn{2}{c|}{Test accuracy}       \\ \hline
\multicolumn{1}{|c|}{Input size} & $\varepsilon$            & \multicolumn{1}{c|}{SGD}                    & DP-SGD  \\ \hline
\multicolumn{1}{|c|}{1000}       & \multirow{5}{*}{0.0355} & \multicolumn{1}{c|}{\multirow{5}{*}{100\%}} & 59.75\% \\ \cline{1-1} \cline{4-4} 
\multicolumn{1}{|c|}{500} &  & \multicolumn{1}{c|}{} & 67.20\%          \\ \cline{1-1} \cline{4-4} 
\multicolumn{1}{|c|}{100} &  & \multicolumn{1}{c|}{} & 84.45\%          \\ \cline{1-1} \cline{4-4} 
\multicolumn{1}{|c|}{50}  &  & \multicolumn{1}{c|}{} & 84.85\%          \\ \cline{1-1} \cline{4-4} 
\multicolumn{1}{|c|}{10}  &  & \multicolumn{1}{c|}{} & \textbf{85.80\%} \\ \hline
\end{tabular}
\vspace{-3pt}
\end{table}
We see that the smaller the input size, the faster the convergence to the true (\sgd) set of parameters (measured by the cosine distance). For the simplest model (input size 10) the distance quickly drops towards 0, indicating that DP training effectively yields a similar parameter vector as non-DP training. By contrast, the cosine distance for the more complex models (i.e. models with input sizes 50, 100, 500, or 1000) decreases at a much slower rate, which highlights that {\bf the DP noise has a significantly detrimental effect on the training process.} Since slower convergence means more training epochs are required (and more epochs means a larger privacy budget) the larger the input size the larger the privacy budget must be to achieve the same model accuracy.
In this case, the prediction task is the same across all models, so there is no benefit from using a larger input size that could potentially offset the detrimental effect of DP noise.

\subsection{Effective Sample Size}\label{sec:eval:simple-size}

To further explore the effect of DP noise, we need to consider the role of the training data size. As with most differential privacy mechanisms, having larger datasets results in better privacy (i.e., a smaller privacy budget $\varepsilon$). This is also the case for \dpsgd{} because the algorithm uses the sampling of mini-batches in each iteration to amplify privacy~\cite{abadi2016deep}. If the batch size is fixed, the larger the dataset the smaller the sampling ratio, which results in lower privacy budget consumed. Therefore, we can reason about the effect of DP noise by varying the size of the training dataset and considering the privacy budget consumed by different model architectures to achieve the same generalization error.

To do this, we use the \mnist{} dataset and train classifiers with \dpsgd{} stopping as soon as the model achieves $85\%$ validation accuracy. We use two FCN models. The first is a ``simple model'' that consists of a single hidden layer with 128 neurons. The second one is a (more) ``complex model'' with 3 hidden layers of 2048, 256, and 64 neurons. We set the same hyperparameters (i.e., batch size, etc.) for both of models, except for the learning rate which we tune optimally in each case. Since \mnist{} has only 60k images, we vary the training dataset size from 10k to 1 million, by random sampling from the data with replacement. To overcome the randomness of the training process and the DP noise, we repeat the experiment 10 times. 

\begin{figure}[!t]
\centering
    \includegraphics[width=0.925\linewidth]{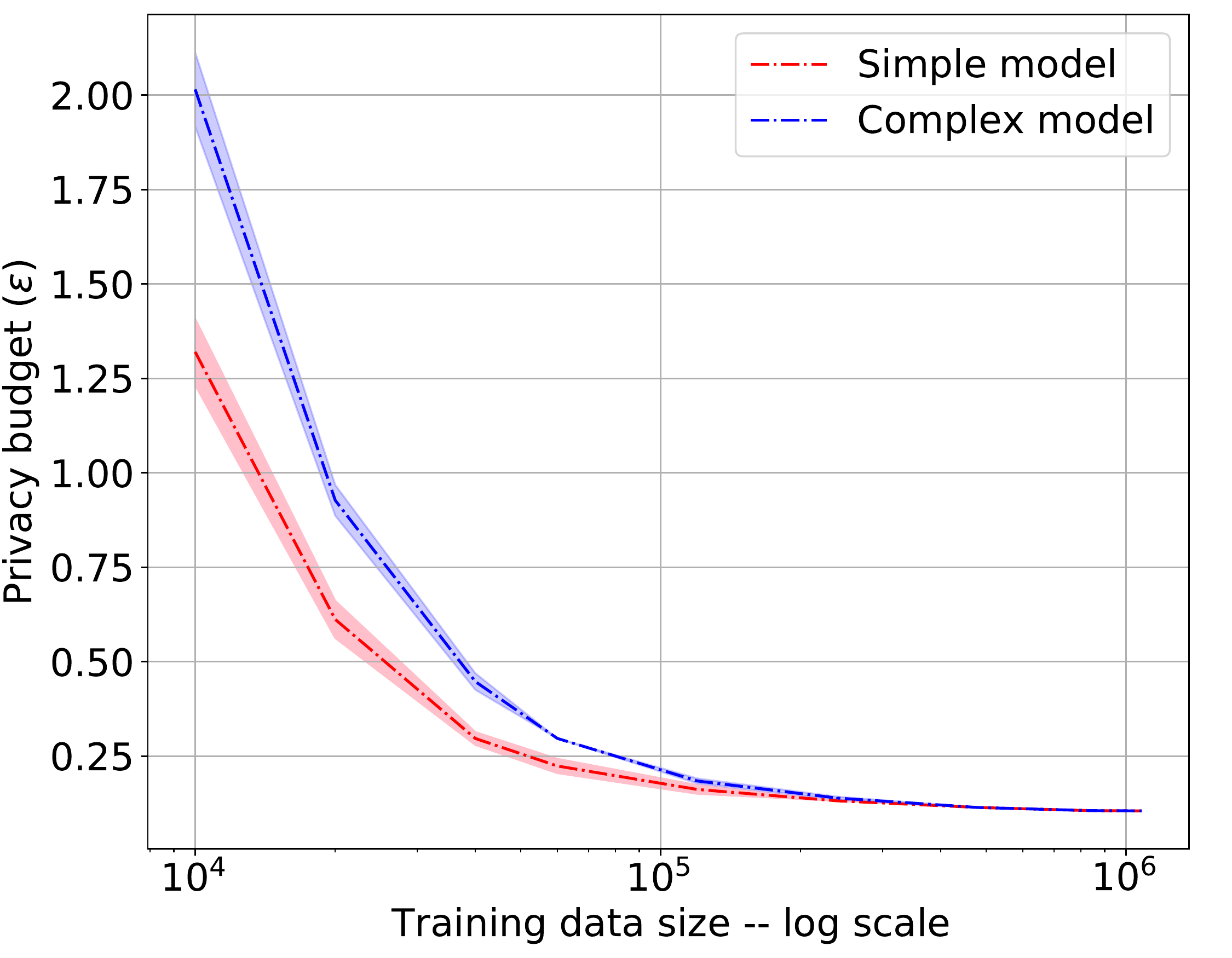}
    \caption{Privacy budget $\varepsilon$ to achieve $85\%$ test accuracy on \mnist{} with \dpsgd{} for varying training data sizes. The shaded area around each curve shows a 99\%-CI.}
    \label{fig:datasize_ci}
    \vspace{-8pt}
\end{figure}
The results are shown in~\cref{fig:datasize_ci} where the x-axis shows the training dataset size (in log scale) and the y-axis shows the privacy budget for training (until stopping at $85\%$ validation accuracy). We see that for a fixed training dataset size, the complex model requires a larger privacy budget $\varepsilon$ to reach the same accuracy. Equivalently, for a fixed privacy budget the complex model requires more training data to achieve the same accuracy. For example for $\varepsilon=1$ it suffices to have $13,574$ training examples for the simple model (to achieve $85\%$ accuracy) whereas it takes $19,145$ training example ($41\%$ more) for the complex model. However, we observe that for large training dataset sizes (e.g., $\geq$ 500k) the two curves converge, which is consistent with related work observations~\cite{tramer2020differentially}.

We found in experiments that the relationship of the privacy budget $\varepsilon$ as a function of the training data size is remarkably predictable. Curve fitting shows that --- for training data sizes large enough to reliably achieve the threshold accuracy --- the relationship is of the form: 
\[ \varepsilon = \ln{\left(\alpha + \frac{\beta}{n}\right)} \ , \]
where $n$ is the training data size and $\alpha \geq 1$ and $\beta > 0$ are a task/dataset dependent parameters. The form of this resembles that of theorems for privacy amplification by subsampling for differential privacy~\cite{balle2020privacy}. Specifically, on the simple model of~\cref{fig:datasize_ci}, we find an almost perfect fit for $\alpha=1.1$ and $\beta=8922.4$.

\subsection{Measuring the Crossover $\varepsilon$} \label{exp:crossover}
Recall from~\cref{sec:framework:crossover} that the crossover epsilon $\varepsilon^\times$ is the privacy budget threshold at which the STW model and PAW model have the same generalization error. Thus, the decision of which architecture to use depends on whether the $\varepsilon$-DP constraint is below the crossover point, i.e.: $\varepsilon \leq \varepsilon^\times$.
\begin{figure}[!t]
\centering
\includegraphics[width=0.925\linewidth]{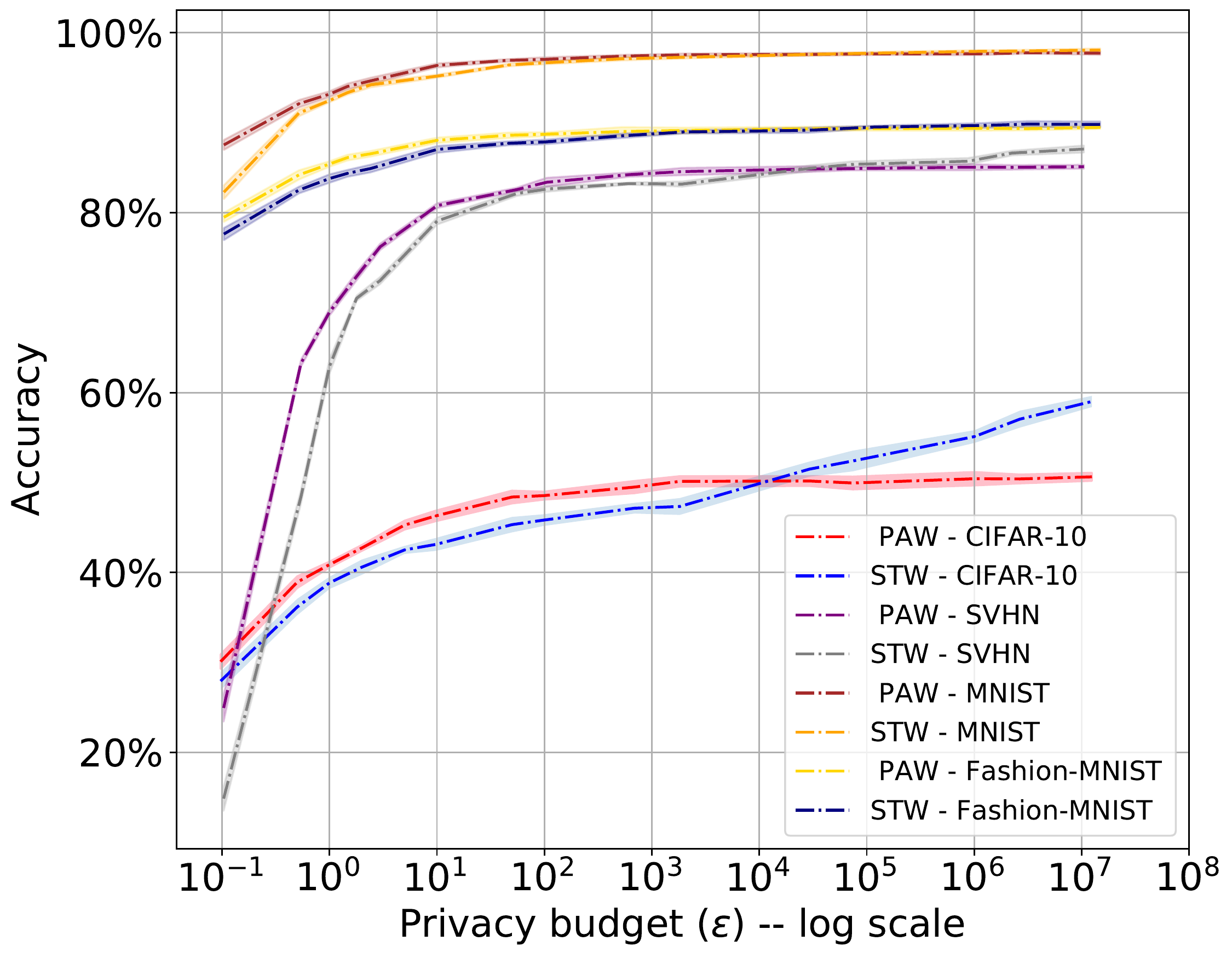}
\caption{Test accuracy on the STW model and PAW model for varying privacy budget $\varepsilon$ on \cifar{}, \svhn{},\fmnist{} and \mnist{} datasets. The shaded area around each curve shows a 99.9\%-CI.}
\label{fig:coeps-multiple}
\end{figure}
\paragraphb{Architecture Search}\label{exp:crossover:arch}
We train FCNs for both STW and PAW with \dpsgd{} on \cifar, \svhn, \mnist, and \fmnist{} for varying $\varepsilon$ from $10^{-1}$ to $10^7$. For each chosen value of $\varepsilon$, we perform a separate grid search in addition to the architecture search to ensure that the learning rate is optimally tuned in each case.\footnote{Unexpectedly, we found that to achieve high accuracy when the privacy budget is large (e.g., $\varepsilon > 10^3$), we need to use large learning rates. We investigated this further and describe our findings in~\cref{exp:svhn-lr}.} 
The results are shown in~\cref{fig:coeps-multiple} where we see that the crossover epsilon $\varepsilon^\times$ is around $10^4$ for these datasets (note: this is so large as to make the privacy guarantee meaningless). The shaded region shows an approximate $99.9\%$-confidence interval on the test accuracy for both the STW model and PAW model. We see that for $\varepsilon$ significantly smaller and significantly larger than the crossover epsilon, the two confidence intervals do not overlap. Interestingly, the largest relative accuracy gap between the PAW and STW models for $\varepsilon \leq \varepsilon^\times$ is not always for very low $\varepsilon$ (i.e., stringent privacy guarantees). For example for \cifar{} it is for values of $\varepsilon$ that are relatively large but significantly lower than the crossover value.

\paragraphb{Feature Selection}\label{exp:crossover:fea}
We estimated the range of $\varepsilon^\times$ on \adult{} and \breast{} taking as the simpler model the one with the subset of features that provides the highest test accuracy model in each case. For \adult{} this gives us six features and for \breast{} a single feature if $\varepsilon$ for both cases is $0.5$. We estimated that the range for the crossover epsilon is  
$[10, 20]$ for \adult{} and $[60,70]$ for \breast{}.
For \adult{}, the range is readily estimated by visual inspection of~\cref{fig:cfs_feature_acc_adult}. For \breast{}, we see from the red line ($\varepsilon = 50$) in~\cref{fig:cfs_feature_acc_breast} that a (slightly) lower accuracy is obtained when using all $43$ features compared to a smaller subset, which implies: $\varepsilon^\times > 50$.

\subsection{Crossover $\varepsilon$ for Model Selection}\label{exp:crossover:model_selection}

\paragraphq{Can FCNs ever outperform CNNs?}
For many image classification tasks, CNNs are known to significantly outperform fully-connected architectures~\cite{lecun1998gradient,urban2016deep}. However, the cost of achieving privacy (even for reasonable $\varepsilon$) can be so large that it reverses this. We use the STW CNN model on \mnist{} but increase the noise level to $2$ and the number of training epochs to $500$. We do this so that both models achieve the exact same $\varepsilon$-differential privacy guarantee ($\varepsilon = 2.11$). When we train this CNN model using \dpsgd, the test accuracy is $93.92\%$. In contrast, the test accuracy of the PAW FCN model (see~\cref{tbl:arch-gensearch}) is $94.94\%$. In other words, {\bf when the privacy budget is limited, it may be better to use a simple FCN model instead of a CNN model}.

\begin{figure}[!t]
\centering
    \includegraphics[width=0.925\linewidth]{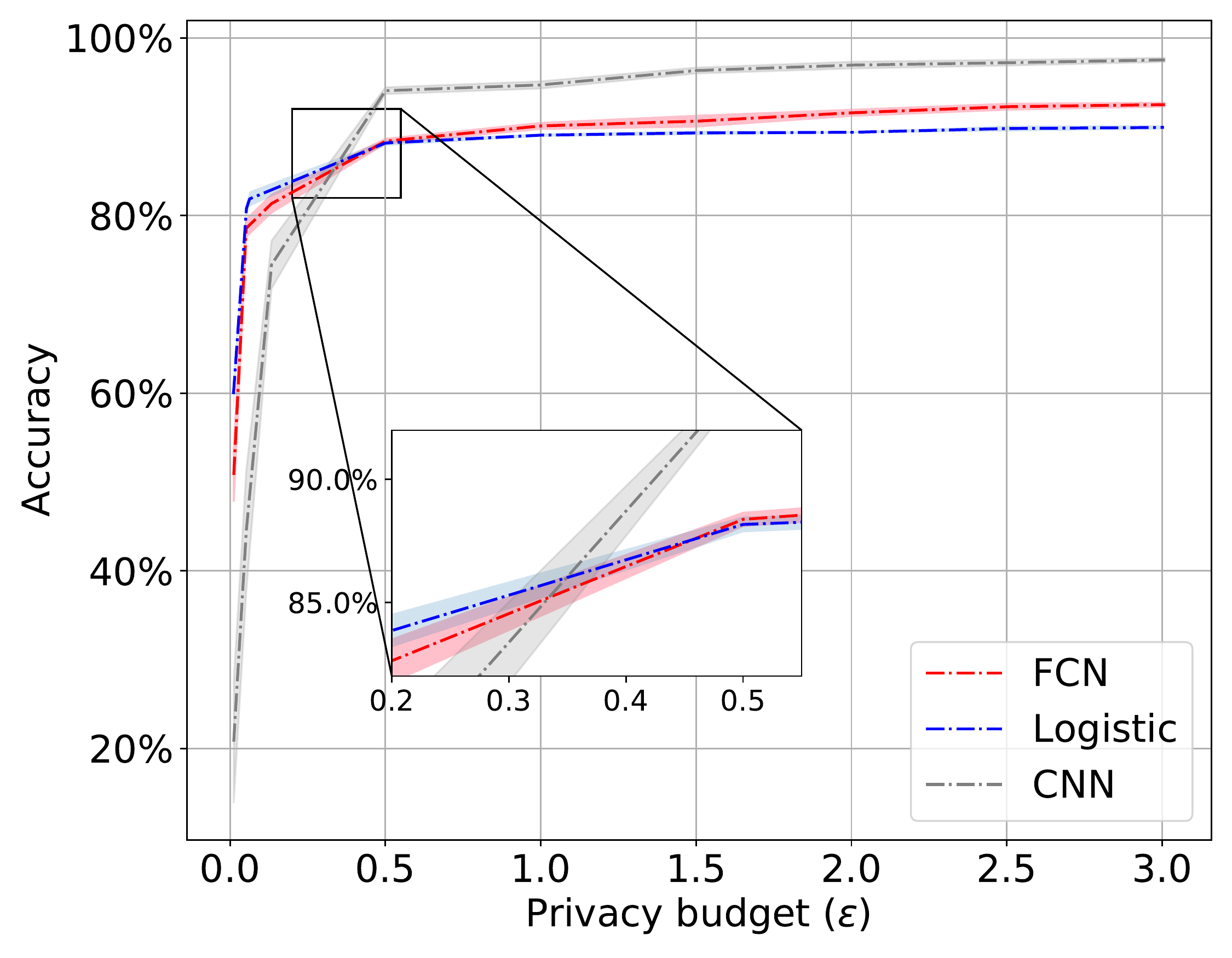}
    \caption{Test accuracy on \mnist{} of Logistic regression model, FCN model and CNN model for varying training $\varepsilon$. The shaded area around each curve shows a 99.9\%-CI.}
    \label{fig:diff_model}
    \vspace{-6pt}
\end{figure}

\paragraphb{Choosing architectures and the crossover $\varepsilon$}
Consider three alternative model architectures for \mnist{}: FCN, CNN, and logistic regression. Clearly, when ignoring privacy CNN is the better choice. However, as we argue in this paper one should use the crossover epsilon to decide. We vary the privacy budget $\varepsilon$ from $0.01$ to $2.0$ and plot the test accuracy of all three model architectures. We show the results in~\cref{fig:diff_model}. (Note that we use a different CNN architecture, i.e., the PAW CNN model, than the previous experiment.) The crossover epsilon $\varepsilon^\times$ is in the range from 0.25 to 0.5. Concretely: if the privacy budget $\varepsilon$ is smaller than $0.25$ then the best choice is the logistic regression model; if the privacy budget $\varepsilon$ is larger than $0.5$ then one should opt for the CNN model. This experiment emphasizes the utility of the crossover epsilon both as a concept and a measurable quantity: {\bf even within a (relatively) small range for the privacy budget $\varepsilon$, the model architecture that yields the highest accuracy can vary widely}.

\subsection{Takeaways} \label{sec:eval:takeaway2}
As we hypothesized, noise added to achieve DP has a detrimental effect on model accuracy, and this detrimental effect increases with model complexity. One can reason about this effect differently by considering the model's training data size and its relationship to the privacy budget. For a fixed privacy budget, one can opt for simplicity or increase the dataset size to enable the use of more complex models. Finally, ML practitioners can measure and use the crossover epsilon ($\varepsilon^\times$) to determine when the benefits from additional model complexity outweighs the detrimental effect of DP noise.

\section{Evaluation: Factors \& Best Practices}\label{sec:exp:factorsflowchat}
We seek to identify factors that contribute to the detrimental effect of DP noise and quantify their effect. In doing so, we discuss the importance of hyperparameter tuning and also compare various architectures to derive some best practices for practitioners.

\subsection{Architectural Factors}\label{sec:exp:factorsflowchat:arch}
Instead of using model architectures with low generalization error when trained with \sgd, we can look to the literature for proven architectures for specific tasks. For example, we can use LeNet~\cite{lecun1998gradient} for \mnist{} 
and compare it to the PAW architecture we found. The results are shown in~\cref{tbl:lenet} where we vary the number of training epochs, batch size, and noise levels. We also use STW model architectures previously found as a baseline. We observe that the performance of the LeNet models is on par with the STW models when training with \sgd, albeit being more parameter efficient using only 61k parameters instead of 197k. However, when training with \dpsgd, the PAW architecture always outperforms both the LeNet and the STW architecture. For example, the PAW model achieves $97.4\%$ accuracy instead of $95.9\%$ for LeNet (150 epochs and batch size of 100). Furthermore, the PAW architecture uses 4736 fewer parameters than LeNet. Taken together, these results suggest that {\bf even architectures that are optimized for a specific task can be outperformed by other (simpler) architectures found in a privacy-aware way}.
\begin{table}[!t]
\centering
\caption{LeNet-5~\cite{lecun1998gradient} on \mnist{} compared with PAW architectures. Our PAW architecture outperforms LeNet under \dpsgd{} despite having fewer parameters.}
\label{tbl:lenet}
\resizebox{\linewidth}{!}{%
\begin{tabular}{cc|c|c|cccc}
\cline{3-4}
 &
   &
  \multicolumn{2}{c|}{Test Accuracy} &
   &
   &
   &
   \\ \hline
\multicolumn{1}{|c|}{Model} &
  Parameters &
  DP-SGD &
  SGD &
  \multicolumn{1}{c|}{Epoch} &
  \multicolumn{1}{c|}{Batch} &
  \multicolumn{1}{c|}{Noise} &
  \multicolumn{1}{c|}{$\varepsilon$} \\ \hline
\multicolumn{1}{|c|}{\cellcolor[HTML]{96FFFB}ST} &
  \cellcolor[HTML]{96FFFB}197130 &
  \cellcolor[HTML]{96FFFB}96.88\% &
  \cellcolor[HTML]{96FFFB}99.42\% &
  \multicolumn{1}{c|}{} &
  \multicolumn{1}{c|}{} &
  \multicolumn{1}{c|}{} &
  \multicolumn{1}{c|}{} \\ \cline{1-4}
\multicolumn{1}{|c|}{\cellcolor[HTML]{96FFFB}PAW} &
  \cellcolor[HTML]{96FFFB}56970 &
  \cellcolor[HTML]{96FFFB}\bf 97.86\% &
  \cellcolor[HTML]{96FFFB}99.38\% &
  \multicolumn{1}{c|}{} &
  \multicolumn{1}{c|}{} &
  \multicolumn{1}{c|}{} &
  \multicolumn{1}{c|}{} \\ \cline{1-4}
\multicolumn{1}{|c|}{\cellcolor[HTML]{96FFFB}LeNet} &
  \cellcolor[HTML]{96FFFB}61706 &
  \cellcolor[HTML]{96FFFB}96.30\% &
  \cellcolor[HTML]{96FFFB}\bf 99.48\% &
  \multicolumn{1}{c|}{\multirow{-3}{*}{300}} &
  \multicolumn{1}{c|}{} &
  \multicolumn{1}{c|}{} &
  \multicolumn{1}{c|}{\multirow{-3}{*}{4.39}} \\ \cline{1-5} \cline{8-8} 
\multicolumn{1}{|c|}{\cellcolor[HTML]{FFFFC7}ST} &
  \cellcolor[HTML]{FFFFC7}197130 &
  \cellcolor[HTML]{FFFFC7}96.61\% &
  \cellcolor[HTML]{FFFFC7}99.34\% &
  \multicolumn{1}{c|}{} &
  \multicolumn{1}{c|}{} &
  \multicolumn{1}{c|}{} &
  \multicolumn{1}{c|}{} \\ \cline{1-4}
\multicolumn{1}{|c|}{\cellcolor[HTML]{FFFFC7}PAW} &
  \cellcolor[HTML]{FFFFC7}56970 &
  \cellcolor[HTML]{FFFFC7}\bf 97.40\% &
  \cellcolor[HTML]{FFFFC7}99.28\% &
  \multicolumn{1}{c|}{} &
  \multicolumn{1}{c|}{} &
  \multicolumn{1}{c|}{} &
  \multicolumn{1}{c|}{} \\ \cline{1-4}
\multicolumn{1}{|c|}{\cellcolor[HTML]{FFFFC7}LeNet} &
  \cellcolor[HTML]{FFFFC7}61706 &
  \cellcolor[HTML]{FFFFC7}95.90\% &
  \cellcolor[HTML]{FFFFC7}\bf 99.36\% &
  \multicolumn{1}{c|}{} &
  \multicolumn{1}{c|}{} &
  \multicolumn{1}{c|}{\multirow{-6}{*}{1}} &
  \multicolumn{1}{c|}{\multirow{-3}{*}{2.98}} \\ \cline{1-4} \cline{7-8} 
\multicolumn{1}{|c|}{\cellcolor[HTML]{FE996B}ST} &
  \cellcolor[HTML]{FE996B}197130 &
  \cellcolor[HTML]{FE996B}93.92\% &
  \cellcolor[HTML]{FE996B}99.34\% &
  \multicolumn{1}{c|}{} &
  \multicolumn{1}{c|}{} &
  \multicolumn{1}{c|}{} &
  \multicolumn{1}{c|}{} \\ \cline{1-4}
\multicolumn{1}{|c|}{\cellcolor[HTML]{FE996B}PAW} &
  \cellcolor[HTML]{FE996B}56970 &
  \cellcolor[HTML]{FE996B}\bf 96.08\% &
  \cellcolor[HTML]{FE996B}99.28\% &
  \multicolumn{1}{c|}{} &
  \multicolumn{1}{c|}{} &
  \multicolumn{1}{c|}{} &
  \multicolumn{1}{c|}{} \\ \cline{1-4}
\multicolumn{1}{|c|}{\cellcolor[HTML]{FE996B}LeNet} &
  \cellcolor[HTML]{FE996B}61706 &
  \cellcolor[HTML]{FE996B}94.30\% &
  \cellcolor[HTML]{FE996B}\bf 99.36\% &
  \multicolumn{1}{c|}{\multirow{-6}{*}{150}} &
  \multicolumn{1}{c|}{\multirow{-9}{*}{100}} &
  \multicolumn{1}{c|}{} &
  \multicolumn{1}{c|}{\multirow{-3}{*}{1.09}} \\ \cline{1-6} \cline{8-8} 
\multicolumn{1}{|c|}{\cellcolor[HTML]{CBCEFB}ST} &
  \cellcolor[HTML]{CBCEFB}197130 &
  \cellcolor[HTML]{CBCEFB}94.92\% &
  \cellcolor[HTML]{CBCEFB}\bf 99.38\% &
  \multicolumn{1}{c|}{} &
  \multicolumn{1}{c|}{} &
  \multicolumn{1}{c|}{} &
  \multicolumn{1}{c|}{} \\ \cline{1-4}
\multicolumn{1}{|c|}{\cellcolor[HTML]{CBCEFB}PAW} &
  \cellcolor[HTML]{CBCEFB}56970 &
  \cellcolor[HTML]{CBCEFB}\bf 94.96\% &
  \cellcolor[HTML]{CBCEFB}99.12\% &
  \multicolumn{1}{c|}{} &
  \multicolumn{1}{c|}{} &
  \multicolumn{1}{c|}{} &
  \multicolumn{1}{c|}{} \\ \cline{1-4}
\multicolumn{1}{|c|}{\cellcolor[HTML]{CBCEFB}LeNet} &
  \cellcolor[HTML]{CBCEFB}61706 &
  \cellcolor[HTML]{CBCEFB}94.72\% &
  \cellcolor[HTML]{CBCEFB}99.32\% &
  \multicolumn{1}{c|}{\multirow{-3}{*}{70}} &
  \multicolumn{1}{c|}{\multirow{-3}{*}{200}} &
  \multicolumn{1}{c|}{\multirow{-6}{*}{2}} &
  \multicolumn{1}{c|}{\multirow{-3}{*}{1.05}} \\ \hline
\end{tabular}
}
\end{table}

\subsection{Feature Engineering}\label{sec:exp:factorsflowchat:fea}
\begin{table}[!t]
\centering
\caption{FCN models on \svhn{} with and without PCA with \sgd{} and \dpsgd{} ($\varepsilon = 2.58$). Due to the cost of privacy, using PCA provides an overall better model.}
\label{tbl:svhn-pca}
\resizebox{0.9\linewidth}{!}{%
\begin{tabular}{ccc|c|c|}

\cline{4-5}
    &  &  & \multicolumn{2}{c|}{Test Accuracy} \\ \hline 
    & \multicolumn{1}{|c}{Workflow} & \multicolumn{1}{|c|}{Parameters} & \dpsgd & \sgd \\ \hline \hline
\multirow{2}{*}{PCA ($k=300$)}    & \multicolumn{1}{|c|}{ST}   & 221,066  & 72.90\% & {\bf 85.32\% } \\ \cline{2-5} 
                                      & \multicolumn{1}{|c|}{PA} & 39,818   & {\bf 76.10\% } & 84.12\% \\ \hline
\multirow{2}{*}{No PCA} & \multicolumn{1}{|c|}{ST}     & 1,577,738 & 64.40\% & {\bf 87.08\% } \\ \cline{2-5} 
                                      & \multicolumn{1}{|c|}{PA} & 55,042   & {\bf 70.38\% } & 86.01\% \\ \hline
\end{tabular}%
}
\vspace{-8pt}
\end{table}

What is the impact on the privacy of dimensionality reduction or feature extraction techniques such as PCA~\cite{abdi2010principal}? To answer this, we train FCN models on \svhn{} with and without PCA for \sgd{} and \dpsgd. For PCA, we only keep the $k=300$ principal components. The parameters for \dpsgd{} are tuned as in~\cref{tbl:arch-gensearch} so that $\varepsilon = 2.58$. \cref{tbl:svhn-pca} shows the results. Due to the cost of achieving privacy, the use of PCA yields a better model. In part, this is because without PCA the number of parameters for the STW model is quite large (1.5M) so the absolute decrease in accuracy in switching from \sgd{} to \dpsgd{} is substantial ($22.62\%$).

PCA was initially leveraged by Abadi et al.~\cite{abadi2016deep} in the paper proposing \dpsgd. However, we show in this paper (as can be seen in~\cref{tbl:svhn-pca}) that whether we use PCA, there exist simpler architectures that provide significantly better accuracy under \dpsgd. Further, observe that the difference in the accuracy of PCA is much larger with \dpsgd{} than with \sgd.

\subsection{Training Factors}\label{sec:exp:factorsflowchat:training_fact}
In cases where the model architecture is fixed (e.g., because only specific architectures are well-suited for a given prediction task) there are several training factors that influence how the phenomenon manifests. In particular we find that tuning learning hyperparameters (i.e., batch size, learning rate, etc.) has a significant effect and one cannot simply rely on the hyperparameter values that work well with \sgd{}. Furthermore, we find that in some cases training only a subset of the network's layers is a viable strategy.

\begin{figure*}[!t]
\centering
\begin{subfigure}{.325\textwidth}
  \centering
  \includegraphics[width=\linewidth]{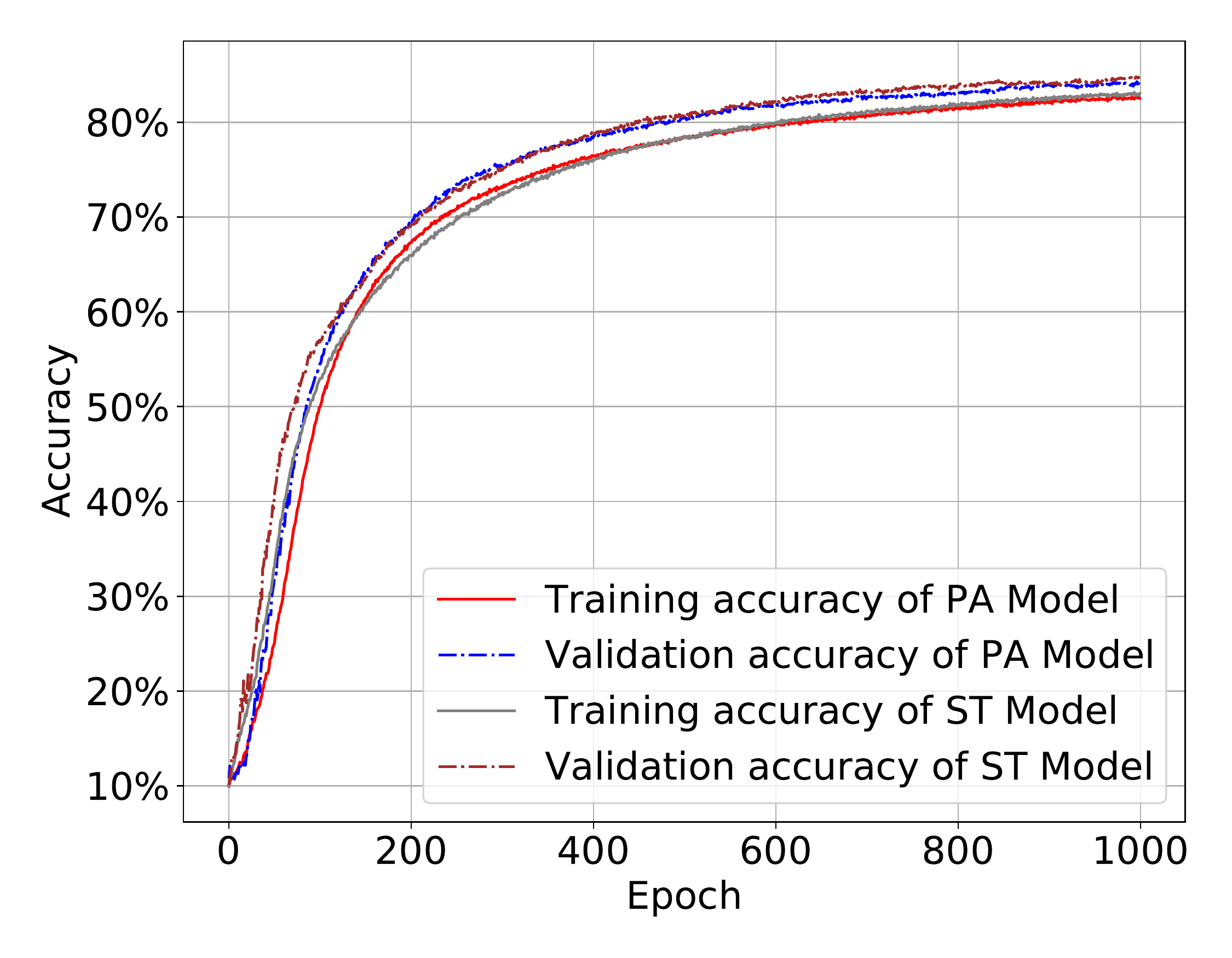}
  \caption{Small learning rate.}
  \label{fig:small_lr}
\end{subfigure}%
\begin{subfigure}{.325\textwidth}
  \centering
    \includegraphics[width=\linewidth]{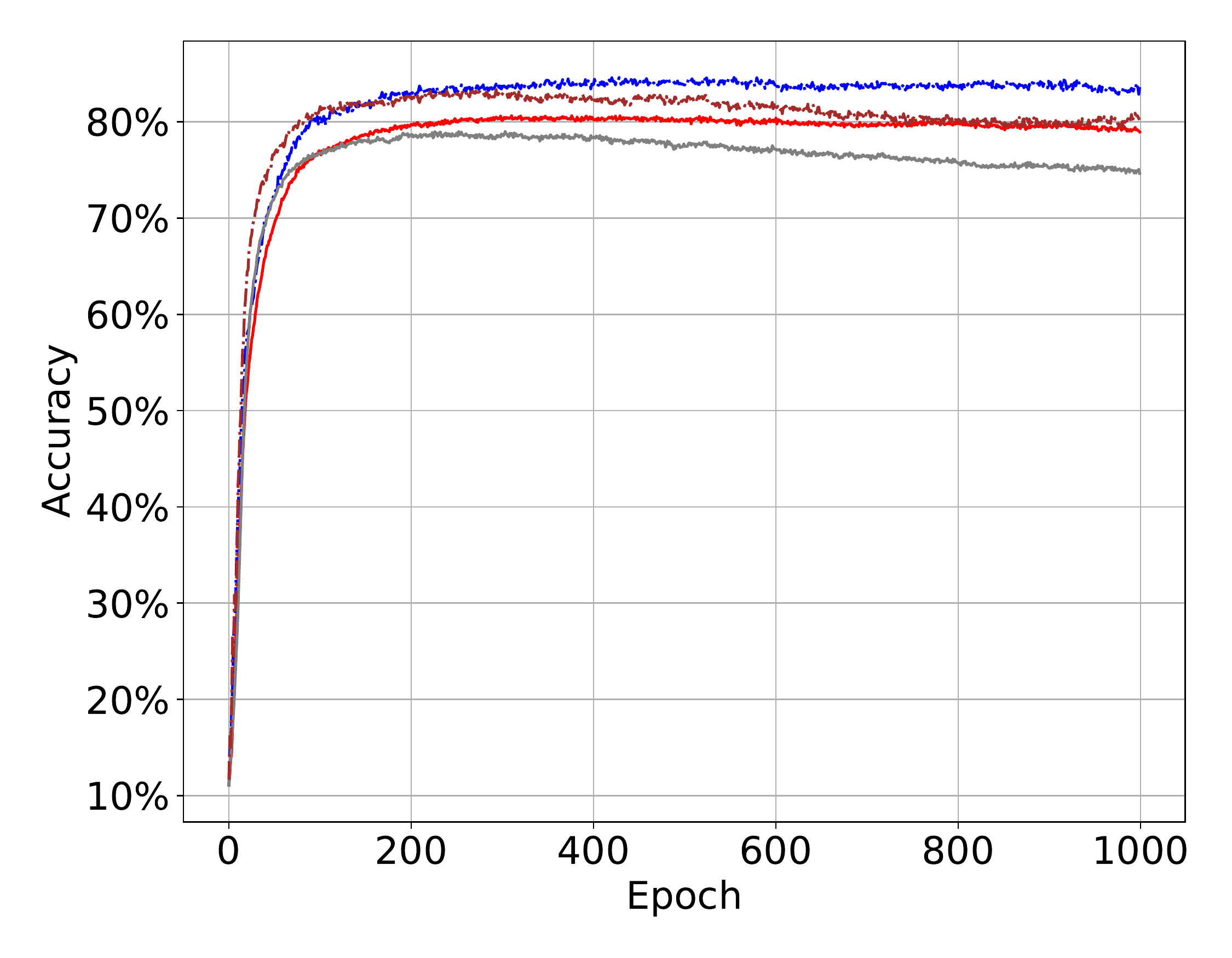}
    \caption{Large learning rate.}
  \label{fig:big_lr}
\end{subfigure}
\begin{subfigure}{.325\textwidth}
  \centering
    \includegraphics[width=\linewidth]{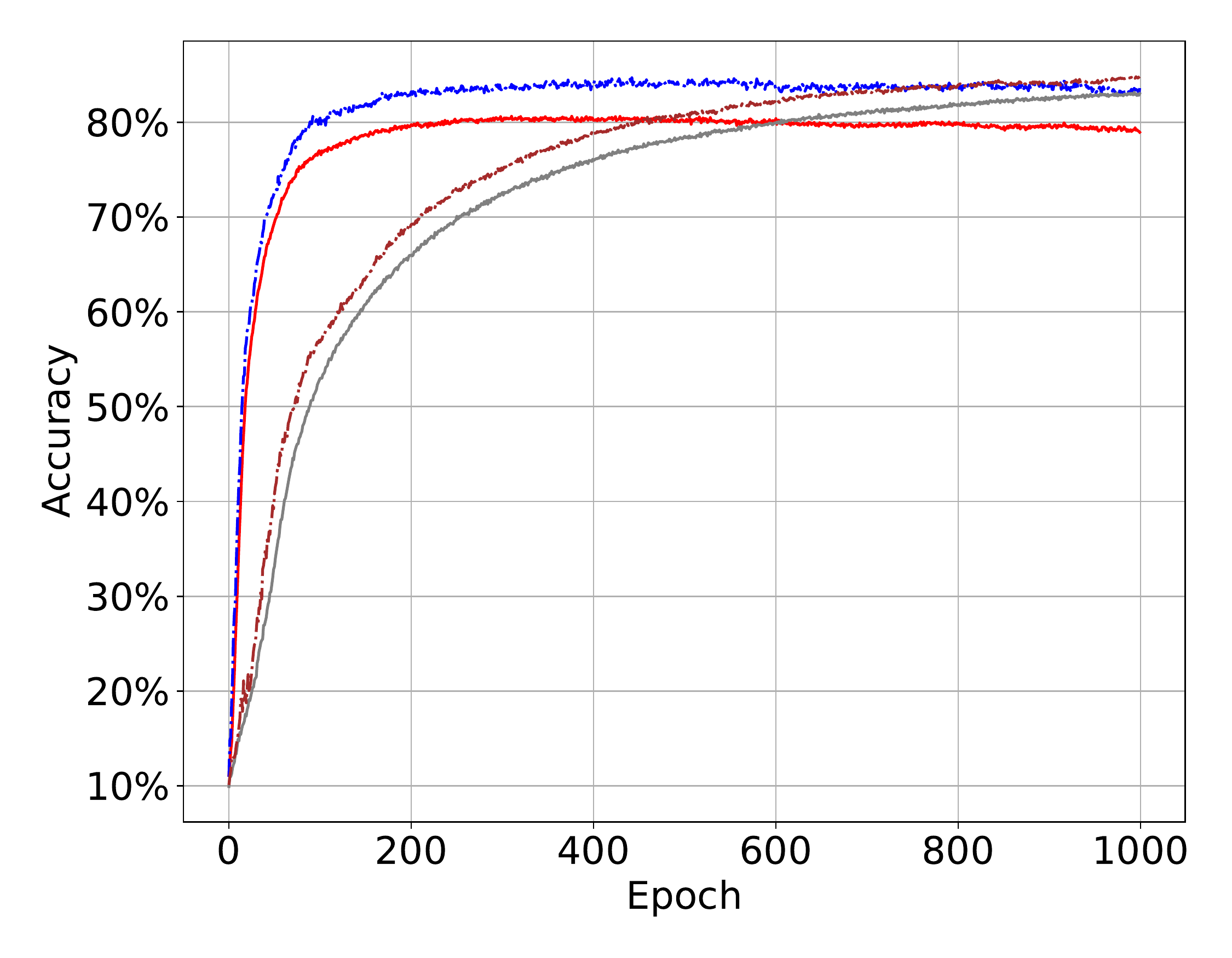}
    \caption{Suitable learning rate in each case. }
  \label{fig:suitable_lr}
\end{subfigure}
\caption{Impact of the learning rate using \svhn{} data. We find that using a close to optimal learning rate in each case (i.e., different for the STW model and PAW model) provides the best accuracy (\cref{fig:suitable_lr}). Using the same learning for both STW and PAW either too small (\cref{fig:small_lr}) or too large (\cref{fig:big_lr}) is misleading and yields worse-than-necessary models.}
\label{fig:lr-study}
\end{figure*}

\paragraphb{Impact of the learning rate}\label{exp:svhn-lr}
Tuning the learning rate is crucial. We found that when using a learning rate that is not suitable, the cost of achieving differential privacy is sometimes hidden. In other words, comparing model architectures and using the same learning rates can be misleading. To show this, we use the \svhn{} dataset and train CNNs with \dpsgd{} for the STW and PAW architectures (for the same $\varepsilon$). This can be seen in~\cref{fig:small_lr,fig:big_lr,fig:suitable_lr}. \cref{fig:small_lr,fig:big_lr} show the behavior when training both STW and PAW models with the same small learning rate (i.e, $0.01$; \cref{fig:small_lr}) and the same larger learning rate (i.e., $0.1$; \cref{fig:big_lr}). In contrast, \cref{fig:suitable_lr} shows the behavior when we train each model with a suitable (i.e., close to optimal) learning rate. Allowing the learning rates to be different between the STW model and the PAW model yields the best-performing model in each case. 

We also observed that when the privacy budget $\varepsilon$ is very large, the model trained with \dpsgd{} outperforms \sgd{}. We discovered that this can be attributed to \dpsgd{}'s use of gradient clipping. We discuss this and other observations about the behavior of \dpsgd{} versus \sgd{} in~\cref{app:clip_gradients}.

\begin{table}[t!]
\centering
\caption{Test accuracy of different training methods. \randomweights{} trains only last 2 layers of the model with \dpsgd{} (other layers are initialized randomly and frozen during training). Normal trains all parameters with \dpsgd{}. For the FCN and CNN models we use \mnist{}. For the RNN model we use \spam{}.}
\label{tab:random_weights_results}
\resizebox{1\linewidth}{!}{%
\begin{tabular}{|c|c|c|c|c|}
\hline
Model                & Params. (trainable)     & Method & Test Accuracy      & $\varepsilon$             \\ \hline
\multirow{4}{*}{FCN} & \multirow{4}{*}{1,362,122 (33,482)} & Normal          & \textbf{90.06\%} & \multirow{2}{*}{2.11}     \\ \cline{3-4}
 &  & RWT    & 84.32\%          &                        \\ \cline{3-5} 
 &  & Normal & 51.34\%          & \multirow{2}{*}{0.341} \\ \cline{3-4}
 &  & RWT    & \textbf{73.06\%} &                        \\ \hline
\multirow{4}{*}{CNN} & \multirow{4}{*}{122,442 (19,328)}  & Normal          & \textbf{97.34\%} & \multirow{2}{*}{5.92}     \\ \cline{3-4}
 &  & RWT    & 96.33\%          &                        \\ \cline{3-5} 
 &  & Normal & 91.20\%          & \multirow{2}{*}{0.341} \\ \cline{3-4}
 &  & RWT    & \textbf{95.20\%} &                        \\ \hline
\multirow{4}{*}{RNN} & \multirow{4}{*}{496,978 (132,610)} & Normal          & \textbf{88.64\%} & \multirow{2}{*}{$1.47 \cdot 10^{11}$} \\ \cline{3-4}
 &  & RWT    & 87.84\%          &                        \\ \cline{3-5} 
 &  & Normal & 79.80\%          & \multirow{2}{*}{0.391} \\ \cline{3-4}
 &  & RWT    & \textbf{87.00\%} &                        \\ \hline
\end{tabular}
}
\end{table}
\paragraphq{Reducing parameters without switching architecture.}
Instead of switching model architecture, a way to effectively reduce the number of parameters is to make some parameters non-trainable (i.e., freeze them after initialization). We call this {\em Random Weights Training} (\randomweights{}). Specifically, we initialize the parameters/weights associated with the first few layers of the model randomly and then freeze them. We then train the model, which effectively tunes only the remaining layers' parameters through backpropagation. 

We evaluate this strategy on FCNs (\cref{table:fcn_random}), CNNs (\cref{tbl:cnn_random}) and RNNs (\cref{tbl:rnn_random}) by only training the last two layers. Results for varying $\varepsilon$ are shown in~\cref{tab:random_weights_results}. {\bf We observe that \randomweights{} achieves higher accuracy than training all layers when the privacy budget is small.} Our results suggest that the reason for this is that the frozen layers act as rudimentary feature extractors and that having less accurate features from those layers is outweighed by the ability to precisely tune the last layers' weights given the larger effective per weight privacy budget. 

Moreover, we discovered that \randomweights{} sometimes allows a complex architecture to outperform a much simpler one. An example of this is shown in~\cref{tab:compara_rand}. Here the complex model has three times as many parameters as the simple model. Based on our other experiments, training the complex model with \dpsgd{} should result in significantly lower accuracy ($49.90\%$) than the simple model ($73.04\%$). However, if we train the complex model with \randomweights{}, it achieves $75.18\%$ test accuracy, outperforming the simple model since fewer parameters are actually trained despite the higher total number of parameters. Thus, it may be more appropriate to think of complexity in terms of the number of {\em trainable} parameters rather than the total number of parameters. This also suggests that a promising area for future research is transfer learning with differential privacy.
\begin{table}[t!]
\centering
\caption{Test accuracy for complex model with \randomweights{} and simple model with normal training.}
\label{tab:compara_rand}
\resizebox{0.8\linewidth}{!}{%
\begin{tabular}{|c|ccc|}
\hline
Model            & \multicolumn{2}{c|}{Complex}                                         & Simple  \\ \hline
Method & \multicolumn{1}{c|}{Normal} & \multicolumn{1}{c|}{\randomweights{}} & Normal \\ \hline
\# of parameters & \multicolumn{1}{c|}{1,362,122} & \multicolumn{1}{c|}{33,482}            & 435,402  \\ \hline
DP accuracy      & \multicolumn{1}{c|}{49.90\%} & \multicolumn{1}{c|}{\textbf{75.18\%}} & 73.04\% \\ \hline
$\varepsilon$    & \multicolumn{3}{c|}{0.34}                                                      \\ \hline
\end{tabular}
}
\end{table}

\begin{figure}[!t]
\centering
\includegraphics[width=0.825\linewidth]{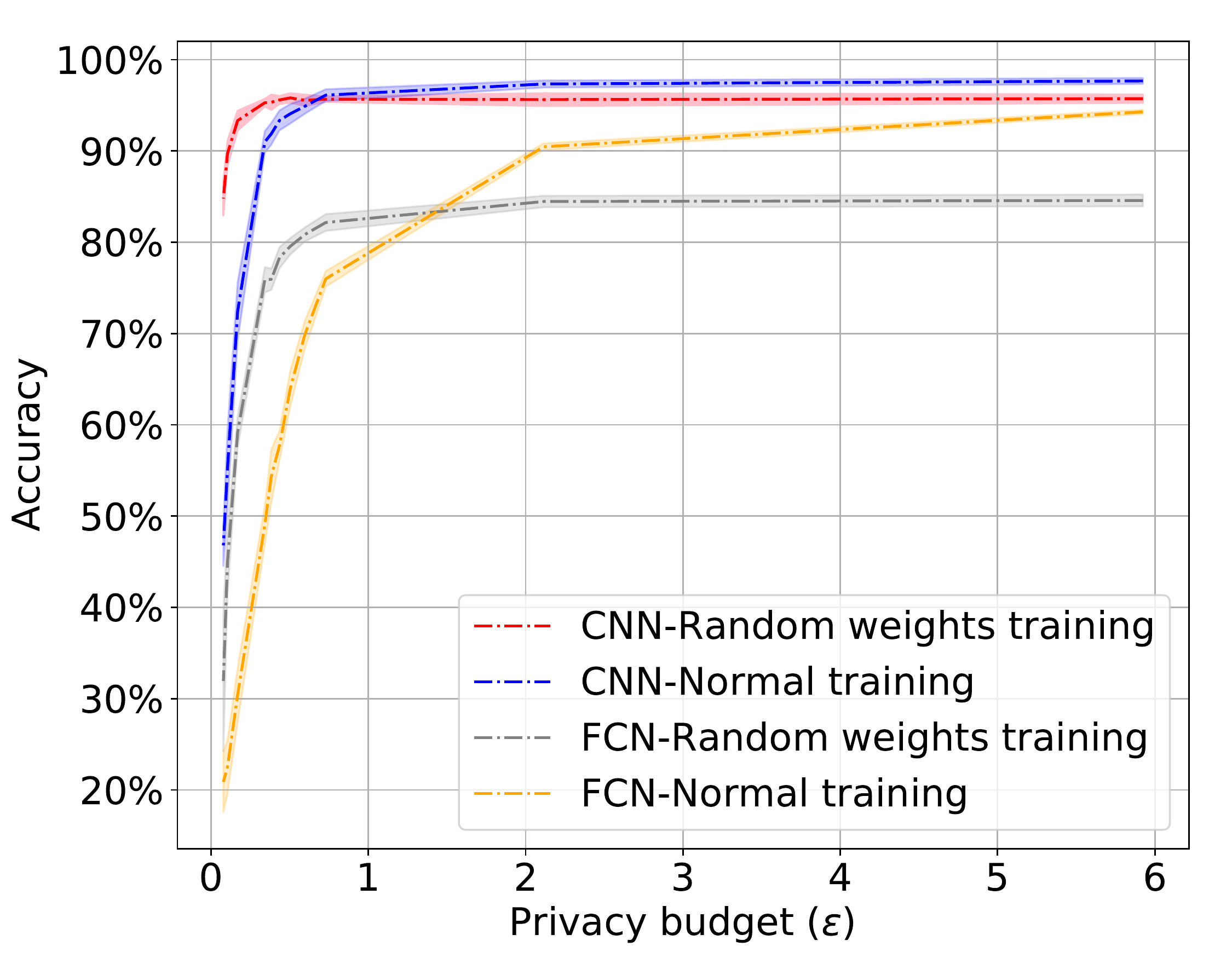}

\caption{Test accuracy on \randomweights{} and normal training with DP-SGD for varying privacy budget $\varepsilon$ on \mnist{} data. It can be seen that the crossover $\varepsilon$ is around 0.6 for CNN model and 1.4 for FCN model. The shaded area around each curve shows a 99.9\%-confidence interval. }
\label{fig:random_crossover}
\end{figure}

The benefits of \randomweights{} are also apparent when we measure the crossover $\varepsilon$. We measure the crossover $\varepsilon$ for FCN and CNN models in~\cref{fig:random_crossover}. We see that whenever the privacy budget is small (i.e., $0.6$ for CNN and $1.4$ for FCN), it is better to use \randomweights{} than train all layers.

\begin{figure*}[ht!]
\centering
\includegraphics[width=0.905\linewidth]{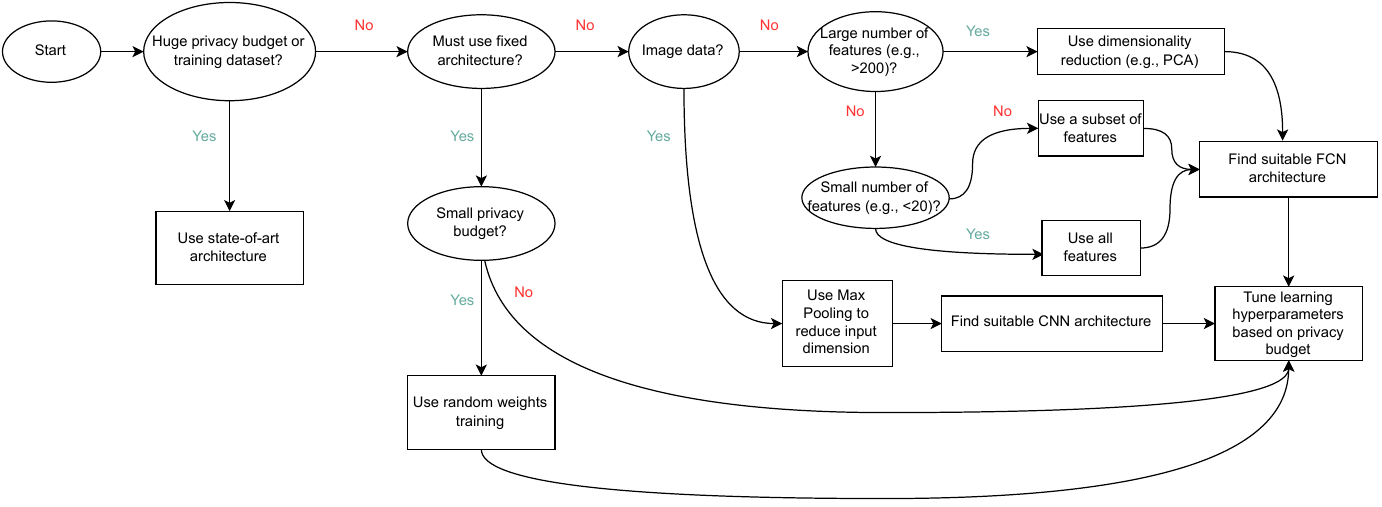}
\caption{Proposed flowchart for differentially-private machine learning. We expect future to refine and expand on these preliminary guidelines.}
\label{fig:flow_chart}
\vspace{-8pt}
\end{figure*}

\subsection{Towards Best Practices}\label{sec:exp:factorsflowchat:flowchart}

Taken together our experimental insights can be distilled down into a set of recommendations or guidelines that we hope can serve as the basis of best practice in the future. For now, we express these recommendations as flowchart (\cref{fig:flow_chart}) that practitioners can use to guide the ML engineering process.

\section{Algorithms}\label{sec:algos}
We propose privacy-aware algorithms for feature selection and architecture selection.

\subsection{Architecture Search}\label{sec:algorithms:arch}
There is significant prior work on neural architecture search~\cite{elsken2019neural}, with some using reinforcement learning~\cite{jaafra2019reinforcement} and others using genetic or evolutionary search methods~\cite{stanley2002evolving,chen2006neuroevolution,xie2017genetic}. The evidence suggests that both reinforcement learning and evolutionary search outperform randomized search~\cite{real2019aging}. 

\paragraphb{PAAS}
We propose a privacy-aware neural network architecture search based on a genetic algorithm. While this algorithm shares a common structure with genetic algorithm based neural architecture search methods, it is designed to take into account the impact of the differential privacy guarantee, and carefully calibrated Laplace noise is added to its fitness function to improve robustness --- and ensure that the search process itself (not just released the model) satisfies differential privacy. We call this algorithm PAAS.

The search space of PAAS primarily includes architectural elements such as the number of layers, activation functions, number of units per layer, etc. However, the search space can be made to also include traditional hyperparameters (e.g., learning rate, batch size, regularization constants, etc.), or those hyperparameters can be optimized separately given a neural network architecture. More precisely, PAAS takes an initial search space of architectures $\modelspace$, which may be thought of as a dictionary mapping architectural choices to a list of possible values. For example:
\vspace{-4pt}
\begin{align*}
    \modelspace = \{ & {\rm num\_layers}: [1, 2, 4, 8, 16], \\
    &{\rm num\_units\_per\_layer}: [64, 128, 256], \\
    & {\rm activation\_fn}: [{\rm TanH}, {\rm Sigmoid}, {\rm ReLU}] \} \ ,
\end{align*}
for a simple neural network search space. Note that the dictionary or list of values for any entry need not be finite.

The full algorithm is described in~\cref{alg:archsearch} and it internally uses an Evolve() function (\cref{alg:evolve}) to produce the next generation's population on the basis of the previous one. The algorithm takes into account privacy through a \train{} function that is an abstraction for the training process, meant to be instantiated in practice with a differentially private training procedure.

Given the search space $\modelspace$, the algorithm initially selects a population of $k$ architectures uniformly at random (line 1). The algorithm then proceeds in $l$ rounds, one for each generation. In each generation, \train{} is invoked on every candidate architecture in the population and the corresponding model is then rank-ordered by fitness value (lines 3 and 4). As stated earlier, the fitness function is the Laplace noised accuracy of the model on the validation data. We found experimentally that this noise increases robustness, but it has the added benefit of making the algorithm differentially private with respect to the validation data. Concretely, we add Laplace noise ${\rm Lap}(\Delta/\varepsilon')$ to the model's validation accuracy with $\Delta=1/n$ where $n$ is the size of the validation dataset (e.g., we set $\varepsilon' = 0.02$ in experiments). The rank-ordered population of architecture (we discard the parameters after ranking) is then used to produce the next generation's population using Evolve(), which does so through a combination of selecting top-scoring architectures, crossover, and mutation (line 6). At the end of the last generation, the best-ranked architecture is selected and returned by the algorithm (lines 9 and 10). As a simple optimization, we use memoization to avoid repeated calls of \train{} for the same architecture during a single run of the algorithm.

For the standard workflow (STW), we use PAAS but instantiate \train{} with a non private learning algorithm (e.g., \sgd).  Once the algorithm terminates and returns the best architecture found, we then train the model using a DP mechanism (e.g., \dpsgd). In contrast, for the privacy-aware workflow (PAW), we use PAAS and instantiate \train{} using a DP mechanism (e.g., \dpsgd{}).

\paragraphb{Randomized Search (\randomsearch)}
As an alternative search strategy to PAAS, we consider Randomized Search (\randomsearch). This strategy is inspired by Appendix D of~\cite{abadi2016deep}. The idea is to randomly and {\em independently} sample a set of $k$ architectures (out of the search space $\modelspace$), evaluate their performance, and pick the one that performs the best. The motivation is that the process can performed to satisfy $\varepsilon$-DP without $\varepsilon$ increasing directly in terms $k$ as it would from naive use of composition results. This is because of a result of Gupta et al.~\cite{gupta2010differentially} (Theorem 10.2). We note here that there is recent work on hyperparameter tuning with differential privacy such as~\cite{papernot2021hyperparameter}.

\paragraphb{Multi-Generation Randomized Search (\genticrandom)}
In practice, PAAS discovers good architectures but its privacy budget increases sharply with the candidate population size and the number of generations due to composition. \randomsearch{} picks architectures randomly so it is an inefficient search strategy, but it achieves differential privacy with a (relatively) small budget. We propose a hybrid search algorithm called {\em Multi-Generation Randomized Search} (\genticrandom) that combines PAAS and \randomsearch{} to obtain the best of both worlds. The basic idea is this: perform a search across multiple generations as in a genetic algorithm but within generations use randomized search. At the end of each generation, we randomly and independently mutate the architecture to create the population of the next generation. Specifically, with some probability $p$ the architecture is mutated randomly and (with probability $1-p$) we re-sample a new architecture uniformly at random from the search space. The initial population is selected through uniformly random sampling of the search space.

\subsection{Feature Selection}\label{sec:algorithms:fs}
Feature selection is the process of selecting the features to use out of a pool of potential features. Typically the goal of feature selection is to improve the performance of the trained model or to reduce the cost of the training process. There exist several feature selection techniques, but in this paper, we consider principal component analysis, max pooling, and correlation-based feature selection.

\paragraphb{Principal Component Analysis (PCA)} 
PCA~\cite{abdi2010principal} is a dimensionality reduction technique that can also be used for feature selection. Given a dataset with $m$ real-valued features, PCA projects the data into a new space defined by a sequence of $m$ principal components that are linearly uncorrelated. By selecting the $k$-first principal components (for $k < m$) the dataset can be expressed such that each data point is represented as $k$ real-valued features (one for each principal component). 

\paragraphb{Max Pooling}
Convolutional neural networks often include down-sampling layers such as average pooling layers and max-pooling layers. A max-pooling layer down-samples its input by taking the maximum pixel value over the filter area for each slice, depth, and channel. We use max pooling for feature selection over raw images that are represented as 3d arrays of pixel values (the 3rd dimension is the channel - RGB). For example, applying (2,2) max pooling to a 128 $\times$ 128 image would result in 64 $\times$ 64 output image, where each pixel value is the max of the corresponding $2\times2$ square of pixel values in the input image. (The operation is applied independently to each channel.)

\begin{algorithm}[!t]
  \caption{Genetic Architecture Search: PAAS} 
  \label{alg:archsearch}
  \begin{algorithmic}[1]
    \Require
      $\modelspace$: search space;
      
      $l$: number of generations;
      
      $k$: size of population (i.e., number of architectures);
    \Ensure
    ${\model^\star}$: the best architecture found
    
    \State $N \gets $ uniformly randomly select $k$ architectures from $\modelspace$
    \For{generation $i$ ($i = 1, 2, \ldots, l$)}
      \State Invoke \train{} on each architecture $a \in N$
      \State Rank order architectures in $N$ by each model's (noised) validation accuracy (i.e., fitness)
      \If {$i < l$}
        \State $N \gets {\rm Evolve}(\modelspace, N)$
        \EndIf
    \EndFor 
    \State Let $\model^\star \gets$ the highest fitness architecture in $N$
    \State \Return{$\model^\star$}
  \end{algorithmic}
\end{algorithm}
\begin{algorithm}[!t]
  \caption{Evolve()} 
  \label{alg:evolve}
  \begin{algorithmic}[1]
    \Require
      $\modelspace$: search space;
      
      $N$: architectures ranked by fitness;
      
      ${\alpha}$: proportion of top-ranking architecture to keep;
      
      ${\beta}$: proportion of random architecture to keep;
      
      $p_{\rm mu}$: probability to randomly mutate an architecture;
       
    \Ensure
    ${N'}$: Next generation's architectures
    
    \State ${N_\alpha} \gets $ ${\alpha}$-proportion top scoring from $N$.
    \State ${N_\beta} \gets $ ${\beta}$-proportion randomly selected from $N \setminus N_t$.
    \State ${N_p} \gets {N_\alpha} \cup {N_\beta}$
    
    \State $N' \gets \emptyset$
    
    \While{$|N'| < |N|$}
        \State Pick architectures $a_1, a_2$ uniformly at random in $N_p$.
        \State $a \gets {\rm crossover}(a_1, a_2)$
        \State $N' \gets N' \cup \{ a \}$
     \EndWhile
    \State For each $a \in N'$, with prob. $p_{\rm mu}$: $a \gets {\rm mutate}_\modelspace(a)$
    \State \Return{$N'$}
  \end{algorithmic}
\end{algorithm}
\paragraphb{Correlation-based Feature Selection (CFS)}
CFS is a technique that is well-suited for datasets with many categorical features~\cite{hall1999correlation}. The goal of CFS is to identify sets of features that are highly predictive of the target class label and at the same time uncorrelated with each other (i.e., we do not want redundant features). Given a dataset with $m$ features, the {\em merit value} of a set $S$ of $k$ features is: %
\begin{align}\label{eq:cfs_merit}
    \merit(S) = \frac{k\ \corr_y(S)}{\sqrt{k + k(k-1) \ \corr(S)}} \ ,
\end{align} \smallskip
where $\corr_y(S)$ is a measure of the correlation of features in $S$ with the class labels $y$ and $\corr(S)$ is a measure of the correlation between all the features in $S$.

Hall~\cite{hall1999correlation} suggests the use of the average symmetrical uncertainty coefficient as a measure of correlation.  
For random variables $x$ and $y$, the symmetrical uncertainty coefficient is defined as: %
\[
    \suc(x, y) = 2 \left( 1 - \frac{H(x,y)}{H(x) + H(y)} \right) \ ,
\] \smallskip
where $H(x)$ (resp. $H(y)$) denotes the entropy of $x$ (resp. of $y$) and $H(x,y)$ denotes the joint entropy of $x$ and $y$.

A challenge of many approaches to feature selection (including CFS) is that given $m$ potential features, there are $2^m - 1$ non-empty possible subsets of features. Even if we fix the number of features $k$ ahead of time, there are $m \choose k$ possible choices. When $m$ is large (e.g., $m \geq 100$), it becomes computationally infeasible to evaluate the merit value of all possible subsets.
\begin{algorithm}[!t]
  \caption{Greedy algorithm for CFS: CFS-Greedy} 
  \label{alg:cfs-greedy}
  \begin{algorithmic}[1]
    \Require
      \ \ ${C}$: set of potential features; \ 
      
      \ ${k}$: desired number of features ($k \leq |C|$);
    \Ensure
      ${S}$: selected features
    
    \State $S \gets \emptyset$
    \While{$|S| < k$}
    \State $s^\star \gets \argmax_{s \in C \setminus S} \merit(S \cup \{ s \})$
    \State $S \gets S \cup \{ s^\star \}$
    \EndWhile
    \State \Return{${S}$}
  \end{algorithmic}
\end{algorithm}

\paragraphb{A greedy algorithm for CFS}
We propose a greedy algorithm to approximately find the best subset of features~(\cref{alg:cfs-greedy}). Given a set of selected features $S$, the algorithm seeks to add whichever feature (not already in $S$) maximizes the merit value~\cref{eq:cfs_merit}. The algorithm stops when the desired number of selected features is reached. (Alternatively, the algorithm could stop whenever adding any features does not increase the merit value.)

\paragraphb{CFS using a genetic search algorithm}
We also propose a genetic algorithm~(\cref{alg:cfs-genetic}). The algorithm uses the merit value as the fitness criterion for each feature subsets (line 3). In each generation, $k$ subsets of features are generated and become part of the population. To generate the next generation, the top$-\alpha$ proportion of subsets by fitness value are kept whereas the others are discarded (line 4). The algorithm then generates the next population through crossover and mutation of the surviving subsets (lines 8 and 9). When it reaches the last generation, the algorithm returns the subset in the current population with the highest merit value (lines 14 and 15). 

\paragraphb{Privacy-aware Feature Selection (PAFS)}
We also propose an alternative way to do feature selection. The idea is to use the set of features $S$ that, when used to train a model with a differentially private process (e.g., \dpsgd) and a specific privacy constraint, provides the best model performance. This is more amenable to privacy than CFS because the goal is to maximize accuracy given differential privacy, not selecting the set of features that has the highest merit score.

However, it is computationally infeasible to try all $2^m - 1$ non-empty possible subsets of features. So, we use a genetic algorithm similar to~\cref{alg:cfs-genetic}, but with the actual performance of the differential private model (e.g., accuracy on the test dataset) as fitness value. We call this algorithm PAFS.

In~\cref{sec:exp:algos:fea}, we evaluate all three ``privacy-aware'' algorithms CFS-Greedy, CFS-GA, and PAFS. Note that the first two are not technically speaking ``privacy-aware'' but since CFS inherently penalizes subsets of features with redundancy, it avoids including features whose costs will likely outweigh their benefits.
\begin{algorithm}[!t]
  \caption{Genetic Algorithm for CFS: CFA-GA} 
  \label{alg:cfs-genetic}
  \begin{algorithmic}[1]
    \Require
      $C$: set of potential features; \ 
      
      $k$: population size; \
      
      $l$: number of generations; \ 
      
      $\alpha$: parent selection proportion; \
      
      $p_{\rm co}, p_{\rm mu}$: crossover and mutate probabilities; \
    \Ensure
      ${S}$: selected set of features
    
    \State Let $P$ be a collection of $k$ randomly selected subsets of features from ${C}$
    \For{generation $i$ ($i = 1, 2, \ldots, l$)}
     \State Rank order each element $S$ of $P$ by $\merit(S)$ value
     \State Take the $\alpha$-top proportion of $P$ as parent set $P'$.
     \State $P \gets \emptyset$
     \While{$|P| < k$}
        \State Select distinct $S_1$, $S_2$ uniformly at random in $P'$.
        \State With prob. $p_{\rm co}$: $Q \gets {\rm crossover}(S_1, S_2)$
        \State With prob. $p_{\rm mu}$: $Q \gets {\rm mutate}(S_1)$
        \State With prob. $1 - p_{\rm co} - p_{\rm mu}$: $Q \gets S_1$
        \State $P \gets P \cup \{ Q \}$ 
     \EndWhile
    \EndFor
    \State Let ${S^\star} \gets \argmax_{S \in P} ( \merit({S}) )$
    \State \Return{$S^\star$}
  \end{algorithmic}
\end{algorithm}
\subsection{Towards Differentially Private Workflows}\label{sec:algos:dpsearch}
Recall from~\cref{sec:framework:problem} the difference between training the model to satisfy ($\varepsilon,\delta$)-DP and ensuring that the entire workflow guarantees differential privacy. In particular, the standard ``drop-in'' workflow does {\em not} guarantee differential privacy even if we train the model with a procedure such as \dpsgd{}. To ensure that a workflow preserves privacy one needs to ensure that no choices are made on the basis of sensitive data except through a differential privacy mechanism. 

If available, public datasets, can be used to alleviate privacy concerns. In such cases, we would still train the final model on the sensitive dataset using differential privacy. In the rest of this section, we analyze the privacy of our feature selection and architecture search algorithms assuming we only have the one sensitive dataset.

\paragraphb{Feature selection algorithms}
Max Pooling incurs no privacy concerns. PCA can be made to satisfy differential privacy by adding noise to the correlation/covariance matrix directly~\cite{dwork2014analyze}. Similarly, we can make our CFS-based feature selection algorithms (CFS-greedy and CFS-genetic) satisfy differential privacy by computing the symmetrical uncertainty coefficients using differential privacy. For this, we can simply add noise to the entropy values themselves as suggested by Bindschaedler et al.~\cite{bindschaedler2017plausible}. Note that in all these cases, the privacy guarantee is independent of the number of different feature sets that the algorithm evaluates because the merit values themselves are just post-processing of the symmetrical uncertainty coefficients.

In contrast, PAFS trains models with differential privacy for each feature set it evaluates. This means that the privacy budget for running PAFS grows with the population size and number of generations. If $k$ feature subsets are considered in each generation and there are $l$ generations, then we need to compose the model training mechanism $c = k l$ times. (Since we use memoization and reuse results of subsets previously used, the total number of composition will be significantly lower than $c$ in most cases. In practice, we can keep track of the number of unique subsets of features tried $c'$ and do the privacy analysis with $c=c'$.) Using advanced composition, this means that if model training satisfies $\varepsilon$-differential privacy then the workflow will satisfy ($\varepsilon',\delta'$)-differential privacy with PAFS for: 
\begin{equation}\label{eq:dpwadvcomp}
     \varepsilon' = \varepsilon \sqrt{2 c \ln{\frac{1}{\delta'}}} + c \varepsilon (e^\varepsilon - 1) \ ,
\end{equation}
where $\delta'$ can be chosen to optimize the tradeoff with $\varepsilon'$.

\paragraphb{Architecture search algorithms}
Similarly for PAAS, if the population size is $k$ at each generation and there are $l$ generations, we need to compose model training $c = k l$ times. This yields the ($\varepsilon',\delta'$)-differential privacy guarantee as~\cref{eq:dpwadvcomp}. Recall that during the search we add noise to the validation accuracy. If the validation data is disjoint from the training data (as it is in our case) we can simply use parallel composition to account for that.
For \randomsearch{} we follow the approach detailed in Appendix D of~\cite{abdi2010principal}. The idea is to set an acceptable loss proportion in terms of the target accuracy which we denote by $x$, take the size of the validation dataset $v$, and the number of architectures evaluated $k$ and then set $\varepsilon' \in (0, \frac{1}{2})$ to ensure that: 
\begin{equation}\label{eq:appd}
 x v \geq \frac{4}{\varepsilon'}  \ln(\frac{1}{\varepsilon'\delta k}) \ ,
\end{equation}
here $\delta$ is not the $\delta$ for differential privacy but a bound on the probability that the target accuracy is not reached. The overall privacy budget of the mechanism is $\varepsilon + 8 \varepsilon'$, where $\varepsilon$ is the privacy budget of training the (final) model.

In terms of access to sensitive data \genticrandom{} behaves like  \randomsearch{} but within each generation, so it suffices to compose sequentially over the generations to compute the total privacy budget. Thus, if we use $l$ generations, the overall privacy budget is: $\sum_{i=1}^{l} \varepsilon_i$, where $\varepsilon_i$ is the privacy budget for generation $i$. We use~\cref{eq:appd} to compute $\varepsilon_i$ where we set $k=k_i$ the number of architectures evaluated in generation $i$ (we keep $x$, $v$, and $\delta$ fixed across generations).

\section{Evaluation: Search Algorithms}\label{sec:exp:algos}
We evaluate the search algorithms proposed in~\cref{sec:algos} both in terms of utility and privacy. Here utility encompasses the quality of the architectures or set of features found and also computational cost.

\subsection{Architecture Search}\label{sec:exp:algos:arch}
\paragraphb{Setup}
For the FCN experiments, the search space $\modelspace$ consists of variation in the number of layers, number of neurons per layer, and activation function (in~\cref{tbl:mlp-searchspace,app:search-space} ). In all cases, we include a dropout layer with dropout rate $0.2$ after each feed-forward layer to mitigate overfitting. We use PCA and keep the first $300$ components in each case.

For the CNN experiments, the search space includes the number of ``basic blocks'' (each basic block contains one convolutional layer followed by one max-pooling layer), the number of filters per convolution layer only for the first basic block (the number of filters for subsequent blocks is set to 2$\times$ the number of previous convolutional layer's filters), the number of fully connected layer's neurons after flattening. This is shown in~\cref{tbl:cnn-search-space} in \cref{app:search-space}. In all cases, the first layer is a max-pooling layer with a pooling size of (2,2) to reduce dimensionality. Kernel sizes for convolutional layers in basic blocks are set to (5,5) for the first convolutional layer and (3,3) for subsequent layers. In all cases, 'ReLU' is the activation function for convolutional layers. After the last convolutional layer, there is a flatten layer followed by a fully-connected layer. This is followed by a dropout layer with a dropout rate of $0.3$ and the final output layer has $10$ units which is the number of classes for all of our image datasets.

For PAAS, we set $=6)$ (number of generations) and $k=10$ (population size). (Our implementation of PAAS is inspired by Matt Harvey's code.\footnote{\url{https://github.com/harvitronix/neural-network-genetic-algorithm}}) We use categorical cross-entropy as the loss function and train each model for $300$ epochs with a batch size of $100$. We selected these values because they provide the best model accuracy on a holdout test set. Since the learning rate is a critical hyperparameter, we searched for close to optimal values in each case. For this, we performed a combination of grid search and manual trial and error. We find that the best learning rates are different not only between the \sgd{} optimizer and the \dpsgd{} optimizer but also for different model architectures. For FCNs, we found learning rates between $0.001$ and $0.0001$ for \sgd{} and $0.1$ to $0.2$ for \dpsgd. For CNNs, we found learning rates around $0.0001$ for \sgd{} and around $0.05$ for \dpsgd{}.

For \randomsearch{} we use the same hyperparameters setting as above and the same search space (\cref{tbl:mlp-searchspace} and \cref{tbl:cnn-search-space}) but we add the trainable option for every layer which means each layer may be set to trainable or not. We set $k=400$.  For \genticrandom{}, we use the same setting as for \randomsearch{} but set $k=40$ for the first generation, $k=20$ for second generation, and $k=10$ for the third and final generation. The probability of mutation is set to $0.7$ and the probability of re-sampling randomly an architecture from the search space is set to $0.3$.

\paragraphb{Results}
In~\cref{sec:algorithms:arch}, we proposed three different privacy-aware architecture search algorithms: PAAS, \randomsearch{} and \genticrandom{}. To compare these algorithms, we use \mnist{} and ensure that the privacy budget to train the model is the same $\varepsilon=2.11$. We measure the model's test accuracy, time to perform the search (measured in GPU hours), and the overall privacy budget of the search (computed as described in~\cref{sec:algos:dpsearch}). We show the results in~\cref{tab:arch_search_compar}.

We find that all three algorithms achieve almost the same model performance because they find similar architectures, but PAAS finds the best model architecture in the shortest time (i.e., less than 15 GPU hours). By contrast \randomsearch{} and \genticrandom{} require approximately 350 and 65 GPU hours respectively to find a comparable model architecture. In terms of privacy, \randomsearch{} has the lowest privacy budget with $\varepsilon = 3.51$, whereas \genticrandom{} and PAAS require a higher privacy budget, i.e.: $\varepsilon = 8.53$ and $\varepsilon = 22.3$, respectively. 

These results highlight the tradeoff between search time and privacy budget. When the privacy for the search is not a concern such as if a non-sensitive or public dataset can be used for the search (the final model would still be trained on sensitive data with \dpsgd), then PAAS finds the best model architecture significantly faster than alternatives. \randomsearch{} finds interesting model architectures (though not the best) with only a small loss in accuracy when the privacy budget is limited, but the search is slow. \genticrandom{} provides a middle ground: it finds an architecture as good as PAAS with lower privacy budget but takes longer.
\begin{table}[t!]
\centering
\caption{Comparison of architecture search algorithms. The privacy budget $\varepsilon$ for training one model is the same in all cases: $\varepsilon = 2.11$.}
\label{tab:arch_search_compar}
\begin{tabular}{l|l|l|l|}
\cline{2-4}
  & Test accuracy & GPU hours & Overall $\varepsilon$ \\ \hline
PAAS      & 94.94\%     & {\bf [10,15]} & 22.3                  \\ \hline
RS        & 95.18\%     & $[340,360]$        & \textbf{3.51}         \\ \hline
MGRS      & \textbf{95.54\% }    & $[60,70]$        & 8.53                   \\ \hline
\end{tabular}
\vspace{-8pt}
\end{table}
\begin{table}[ht!]
\centering
\caption{STW and PAW architectures for \randomsearch{} with \randomweights{}.}
\vspace{-4pt}
\label{tab:random_seach}
\resizebox{1\linewidth}{!}{%
\begin{tabular}{cc|cc|cc|c}
\cline{3-6}
 &
   &
  \multicolumn{2}{c|}{Test Accuracy} &
  \multicolumn{2}{c|}{Gap} &
   \\ \hline
\multicolumn{1}{|c|}{Workflow} &
  Total Params (Trainable) &
  \multicolumn{1}{c|}{DP-SGD} &
  SGD &
  \multicolumn{1}{c|}{DP-SGD} &
  SGD &
  \multicolumn{1}{c|}{$\varepsilon$} \\ \hline
\rowcolor[HTML]{96FFFB} 
\multicolumn{1}{|c|}{\cellcolor[HTML]{96FFFB}STW} &
  674,762 (658,314) &
  \multicolumn{1}{c|}{\cellcolor[HTML]{96FFFB}92.04\%} &
  \textbf{97.97\%} &
  \multicolumn{1}{c|}{\cellcolor[HTML]{96FFFB}} &
  \cellcolor[HTML]{96FFFB} &
  \multicolumn{1}{c|}{\cellcolor[HTML]{96FFFB}} \\ \cline{1-4}
\rowcolor[HTML]{96FFFB} 
\multicolumn{1}{|c|}{\cellcolor[HTML]{96FFFB}PAW} &
  9,610 (9,610) &
  \multicolumn{1}{c|}{\cellcolor[HTML]{96FFFB}\textbf{95.18\%}} &
  96.84\% &
  \multicolumn{1}{c|}{\multirow{-2}{*}{\cellcolor[HTML]{96FFFB}3.14\%}} &
  \multirow{-2}{*}{\cellcolor[HTML]{96FFFB}-1.13\%} &
  \multicolumn{1}{c|}{\multirow{-2}{*}{\cellcolor[HTML]{96FFFB}2.11}} \\ \hline
\rowcolor[HTML]{FFCCC9} 
\multicolumn{1}{|c|}{\cellcolor[HTML]{FFCCC9}STW} &
  674,762 (658,314) &
  \multicolumn{1}{c|}{\cellcolor[HTML]{FFCCC9}72.28\%} &
  \textbf{97.97\%} &
  \multicolumn{1}{c|}{\cellcolor[HTML]{FFCCC9}} &
  \cellcolor[HTML]{FFCCC9} &
  \multicolumn{1}{c|}{\cellcolor[HTML]{FFCCC9}} \\ \cline{1-4}
\rowcolor[HTML]{FFCCC9} 
\multicolumn{1}{|c|}{\cellcolor[HTML]{FFCCC9}PAW} &
  76,810 (10,250) &
  \multicolumn{1}{c|}{\cellcolor[HTML]{FFCCC9}\textbf{91.61\%}} &
  92.70\% &
  \multicolumn{1}{c|}{\multirow{-2}{*}{\cellcolor[HTML]{FFCCC9}19.33\%}} &
  \multirow{-2}{*}{\cellcolor[HTML]{FFCCC9}-5.27\%} &
  \multicolumn{1}{c|}{\multirow{-2}{*}{\cellcolor[HTML]{FFCCC9}0.34}} \\ \hline
\end{tabular}
}
\end{table}
\paragraphb{Architecture search with \randomweights{}}
In our experiments, we found out that using \randomweights{} with architecture search results in some interesting architectures that are well-suited for stringent privacy budget settings. \cref{tab:random_seach} shows a comparison of PAW and STW architectures trained with RWT. The PAW model found by \randomsearch{} when training $\varepsilon = 0.34$ only experiences a small decrease in accuracy (1.09\%) when using \dpsgd{} for training compared to \sgd{}.

\subsection{Feature Selection}\label{sec:exp:algos:fea}
We compare PAFS, CFS-Greedy, and CFS-GA experimentally using the \breast{} dataset with objective perturbation. In addition, we include selecting random features (``Random'') as a baseline and also evaluate the standard workflow (i.e., selecting ``all'' available features). PAFS chooses the number of features to use automatically, and so does CFS-GA. (Our implementation of PAFS is inspired by the DEAP framework.\footnote{\url{https://github.com/kaushalshetty/FeatureSelectionGA}}) In contrast,  CFS-Greedy and Random require a target size. We set this size to be the same as the size of the CFS-GA selected subset.

The results are shown in~\cref{tbl:fs_comp_breast}. Consistent with our previous results, it can be seen that for small $\varepsilon$, all of the methods evaluated outperform the standard workflow (``All'') which selects all of the available features. Surprisingly perhaps, even a random selection of features (``Random'') outperforms it. This is because, for small $\varepsilon$, many features incur a cost (due to privacy) that is greater than their predictive benefit. Thus, in some cases {\bf using fewer features is better even if those features are selected at random}.

Overall, we find that PAFS outperforms all other methods because it is able to take into account the true cost of achieving privacy and can tune the subset size to $\varepsilon$. Other methods do not taken into account the privacy budget.
\begin{table}[!t]
\centering
\caption{Comparison of test accuracy of feature selection algorithms using \breast{} for varying $\varepsilon$. 'All' is the STW model. PAFS, CFS-Greedy, and CFS-GA are our proposed privacy-aware algorithms. 'Random' is a baseline that selects features at random. PAFS yields significantly better models. Note that 'All' is outperformed by 'Random' in many cases.}
\label{tbl:fs_comp_breast}
\smallskip
\resizebox{\linewidth}{!}{%
\begin{tabular}{l|l||l|l|l|l|} \hline
$\varepsilon$ & All & PAFS & CFS-Greedy & CFS-GA  & Random \\ \hline
0.01 & 50.11\% & 49.99\%          & 50.30\% & \bf 50.44\%  & 49.84\% \\
0.1  & 50.98\% & \textbf{52.09\%} & 51.27\% & 51.72\%         & 51.71\% \\
0.5  & 53.75\% & \textbf{54.92\%} & 54.05\% & 54.49\%         & 53.34\% \\
1    & 53.85\% & 55.12\%          & 55.20\% & \bf 55.38\%  & 54.23\% \\
1.5  & 55.34\% & 55.78\%          & 55.63\% & \bf 55.82\%  & 55.35\% \\
2    & 55.43\% & \textbf{56.84\%} & 56.21\% & 56.47\%  & 55.67\% \\
5    & 56.74\% & \textbf{64.21\%} & 60.23\% & 62.34\%  & 59.30\% \\
10   & 61.78\% & \textbf{68.75\%} & 67.09\% & 67.41\%    & 66.87\% \\
20   & 68.13\% & \textbf{71.00\%} & 70.36\% & 69.53\%         & 70.12\% \\
25   & 69.59\% & \textbf{73.48\%} & 70.59\% & 69.74\%          & 70.61\% \\
30   & 70.75\% & \textbf{71.52\%} & 70.62\% & 69.85\%          & 70.67\% \\
35   & 71.45\% & \textbf{71.72\%} & 70.70\% & 69.92\%         & 70.66\% \\
40   & 72.11\% & \textbf{73.60\%} & 70.71\% & 69.94\%          & 70.57\%\\
\hline
\end{tabular}%
} 
\end{table}

\paragraphb{Privacy analysis}
For PAFS, we use $l=10$ generations and set the population size to $k=600$. We use $\varepsilon=0.1$ for each feature set, which means the total privacy budget is $49.63$. For CFS-Greedy and CFS-GA the privacy budget is: $2.31$. What is important to notice here is that for CFS-Greedy and CFS-GA the privacy budget does not include training the final model. Nevertheless, if privacy of feature selection is a concern then CFS-Greedy and CFS-GA are good choices.

\paragraphb{Running time}
PAFS is not computationally feasible to use in all scenarios because of the large computational cost of training the model. When the number of possible features is large, finding good solutions with PAFS involves training a very large number of models with differential privacy. For example, PAFS with objective perturbation~\cite{chaudhuri2011differentially} (for $k=600$ and $l=10$) takes about 9 hours to run on our hardware because it involves training about $2300$ models for unique feature subsets. By contrast, the running time is (only) 3.8 seconds for CFS-Greedy and 57.6 seconds for CFS-GA (with $k=5000$ and $l=10$).

\section{Related Work}\label{sec:related}
There is a growing body of research on differentially private machine learning models training. The most prominent mechanism is \dpsgd{}~\cite{abadi2016deep}, which recent work such as~\cite{nasr2020improving,zhou2020bypassing,papernot2020tempered,davody2020robust} attempts to improve in various ways. Some examples include: refining how the gradient clipping process deals with bias \cite{xu2020removing,chen2020understanding}, projecting/encoding into a subspace \cite{zhou2020bypassing,nasr2020improving}, using tempered sigmoid activation functions \cite{papernot2020tempered}, batch size normalization~\cite{davody2020robust}, or simply an increased batch size \cite{bagdasaryan2019differential,van2018three,papernot2019making}.

Aside from \dpsgd{}, there are a few other techniques to learn models with differential privacy. For example, Zhang et al. propose PrivGene~\cite{zhang2013privgene} which uses a genetic algorithm to find the best parameters for several simple ML models, which they make differentially private using their enhanced exponential mechanism. Whereas PrivGene leverages genetic algorithms to tune parameters, we use genetic algorithms to search for novel neural network architectures and feature subsets.
There has also been work on differentially private feature selection such as~\cite{thakurta2013differentially,yang2014differentially,anandan2018differentially}.

Despite efforts to make training differentially private models practical, several researchers have pointed out the large utility and accuracy cost of existing techniques~\cite{bagdasaryan2019differential,jayaraman2019evaluating}.
Avent et al.~\cite{avent2020automatic} attempt to alleviate this cost by optimizing hyperparameters for privacy and utility by defining first a Pareto front for the DP algorithm. Tram\`er et al.~\cite{tramer2020differentially} improve performance by using handcrafted Scatternet features, although this causes the model to perform differently than CNN or similar models. 
Papernot et al.~\cite{papernot2019making}, and similar papers~\cite{zhang2013privgene,abadi2016deep,papernot2020tempered}, find that by selecting hyperparameters, optimizers, and activation functions specifically for \dpsgd, model accuracy can be improved. Luo et al.~\cite{luo2021scalable} propose the use of transfer learning on public data to facilitate training with fewer parameters, thereby reducing the loss of utility. Finally, some works~\cite{morsbach2021architecture,cheng2021dpnas} observe that different architectures perform differently at different privacy levels, and Chen et al.~\cite{cheng2021dpnas} propose an algorithm to take this into account.

We believe prior works observed manifestations of the phenomenon we study in this paper but have overlooked its significance. Our paper seeks to explicate this phenomenon, performs experiments to quantify it, and distill our insights into actionable takeaways for practitioners. We show that the phenomenon is {\em not} an oddity limited to the tweaking of hyperparameters such as the learning rate or the activation function with \dpsgd{}, but that it instead (potentially) manifests in every aspect of the modern machine learning workflow. Further, we propose ``privacy-aware'' algorithms that can account for the effect of DP noise.

\section{Discussion and Future Directions}\label{sec:discussion}
A crucial point to contextualize our results and their broader implications is that there are two desirable but orthogonal properties properties of a learning algorithm and (or) workflow: (1) {\em privacy awareness} and (2) {\em differential privacy}. Informally, differential privacy has to do with what the algorithm leaks about the input dataset. By contrast, privacy awareness has to do with whether choices made by the algorithm are influenced by the anticipated cost of achieving privacy. For example, the feature selection algorithm that always selects all available features without looking at the data does satisfy differential privacy, but it is not privacy aware. The consequence of this observation is that we cannot solve the problem by simply combining DP mechanisms, one for each step of the workflow (e.g., DP dataset preprocessing, DP feature selection, DP hyperparameter selection, DP model training, etc.). We need privacy-aware versions of these DP mechanisms.

\paragraphb{Future directions}
Our results call attention to several directions for future research. Perhaps the most obvious is the design of better privacy-aware algorithms and the design of privacy-aware algorithms for other steps of the workflow (e.g., feature engineering, data augmentation). Another direction is to extend the scope of consideration for privacy-aware algorithms by considering other types of models (e.g., decision trees, ensembles, etc.) or applications beyond supervised learning (e.g., clustering, dimensionality reduction) or machine learning. Finally the empirical effectiveness of our random weights training technique motivates the future investigation into transfer learning with differential privacy.

\bibliographystyle{IEEEtran}
\bibliography{refs}

\appendix
\appendices

\section{Proofs}\label{app:proofs}
In this section we provide a proof of~\cref{lem:gaussiannoiselinearmodel}.
\begin{proof}[Proof of~\cref{lem:gaussiannoiselinearmodel}]
    Let $\tilde{\params}$ and $\tilde{\params'}$ as in the Lemma. Let $({\bf x}, y)$ be any data point such that ${\bf x} \neq 0$, we have that:
    \begin{align*}
        {\rm err}(\tilde{\params}, {\bf x}, y) &= \left(y - \sum_{i=1}^{m} (\theta_i + Z_i) x_i \right)^2 \\
        &= \left(c - \sum_{i=1}^{m} Z_i x_i \right)^2 \ , 
    \end{align*}
    where we write $c = y - \sum_{i=1}^{m} \theta_i x_i$, which is the difference between the true label and the model's prediction on ${\bf x}$. Note that: $c^2 = {\rm err}(\params, {\bf x}, y)$.
    Since $c$ is constant and each $Z_i$ has mean $0$, we have that the expectation of the error is:
    \begin{equation}\label{eq:err_dp_cmodel}
        \mathbb{E}[ {\rm err}(\tilde{\params}, {\bf x}, y)] = c + \mathbb{E}[(\sum_{i=1}^{m} Z_i x_i)^2] \ .
    \end{equation}
    Each summand $x_i Z_i$ in~\eqref{eq:err_dp_cmodel} is a Gaussian random variable with mean 0 and variance $x_i^2 \sigma^2$. The inner sum then becomes the sum of $m$ independent Gaussian random variable with mean 0 but with variances $a_i = x_i^2 \sigma^2$. Since the sum of independent Gaussians RVs is also Gaussian we can rewrite~\eqref{eq:err_dp_cmodel} as: \[ \mathbb{E}[{\rm err}(\tilde{\params}, {\bf x}, y)] =  c + \mathbb{E}[W^2] , \]
where $W \sim \mathcal{N}(0, \sigma^2 ||{\bf x}||^2_2)$. Writing $a = \sigma ||x||_2$, we see that $W^2 / a^2$ follows a Chi-square distribution with 1 degree of freedom so that $\mathbb{E}[W^2] = a^2$.

    For the error of $\model'$ we have:
    \begin{align*}
        {\rm err}(\tilde{\params'}, {\bf x}, y) &= \left(c + \theta_m x_m - \sum_{i=1}^{m-1} Z_i' x_i \right)^2 
    \end{align*}
    Again since $Z_i'$ has mean 0, expanding the following expression and taking the expectation yields:
        \begin{align*}
        \mathbb{E}[ {\rm err}(\tilde{\params'}, {\bf x}, y)] &= c^2 + 2 c \theta_m x_m\\
        &+ (\theta_m x_m)^2 + \mathbb{E}[(\sum_{i=1}^{m-1} Z_i' x_i)^2] \ . \nonumber
    \end{align*}
    As before, we recognize that $\sum_{i=1}^{m-1} Z_i' x_i$ is a sum of $m-1$ independent Gaussians RVs with mean 0 and variance $b_i = x_i^2 \sigma'^2$ so that we let $W'\sim \mathcal{N}(0, \sigma'^2 ||{\bf x'}||^2_2)$. Writing $b = \sigma' ||{\bf x'}||_2$, we have that $W'^2 / b^2$ follows a Chi-square distribution with 1 degree of freedom.
    
    Putting the above together we have that the expected squared error under $\model'$ is less than or equal that under $\model$ whenever:
    \begin{align}\label{eq:err_dpmodel_quadratic}
        (\theta_m x_m)^2 + 2 c \theta_m x_m &\leq \mathbb{E}[W^2] - \mathbb{E}[W'^2] \\
        &= a^2 - b^2 \ . \nonumber
    \end{align}
    \cref{eq:err_dpmodel_quadratic} is a quadratic inequality in terms of $\theta_m$. If $x_m = 0$ then the (strict) inequality can only be satisfied provided that $a^2 > b^2$ which implies that $\sigma' < \sigma$ since in that case $||{\bf x}||_2 = ||{\bf x'}||_2$. Otherwise (if $x_m \neq 0$) we can solve the inequality for $u = \theta_m x_m$, in which case the condition for \cref{eq:err_dpmodel_quadratic} to be satisfied is:
    \[
        -\sqrt{c^2 + a^2 - b^2} - c \leq u \leq \sqrt{c^2 + a^2 - b^2} - c \ .
    \]
    Since $a^2 \geq b^2$, we have that $\sqrt{c^2 + a^2 - b^2} \geq |c|$. Applying this on $|u| = |\theta_m| |x_m|$, it follows that:
    \[
        |\theta_m| \leq \frac{\sqrt{c^2 + a^2 - b^2} - |c|}{|x_m|} \ . \qedhere
    \]
\end{proof}

\section{Search space} \label{app:search-space}
In this section, we show the details of the search space (\cref{tbl:mlp-searchspace} and \cref{tbl:cnn-search-space}) for the architecture search experiments.

\begin{table}[!h]
    \centering
    \caption{Architecture search space for FCNs.} \label{tbl:mlp-searchspace}
        \resizebox{0.95\linewidth}{!}{%
        \begin{tabular}{c|c|}
            \cline{2-2}
            & Possible Values \\
            \hline \hline
            Number of layers  & \{ 1, 2, 3 \} \\
            \hline
            First layer's neurons & \{ 64, 128, 512, 1024, 2048  \} \\
            \hline
            Second layer's neurons &  \{ 64, 128, 256  \} \\
            \hline
          Third layer's neurons & \{ 10, 16, 32, 64  \}  \\
            \hline
            First layer's activation & \{ ReLU, Sigmoid, TanH  \} \\
            \hline
            Second layer's activation & \{ ReLU, Sigmoid, TanH \} \\
            \hline
            Third layer's activation & \{ ReLU, Sigmoid, TanH \} \\
            \hline
        \end{tabular}  
        } %
\end{table}
\begin{table}[h!]
    \centering
    \caption{Architecture search space for CNNs.} \label{tbl:cnn-search-space}
        \resizebox{0.95\linewidth}{!}{%
        \begin{tabular}{c|c|}
            \cline{2-2}
            & Possible Values \\
            \hline \hline
            
            \hline
            Number of basic blocks & \{ 2, 3 \} \\
            \hline
            Filters of basic block 1 & \{ 16, 32, 48 \} \\
            \hline
            Units of fully-connected layer & \{ 32, 64, 128, 512 \} \\
            \hline
           
        \end{tabular}
         } %
\end{table}

\section{Additional Experiments} \label{app:exp}
In this section, we provide additional experimental results.

\subsection{What about RNNs?} \label{app:exp:RNN}
We run experiments on the \spam{} dataset with a simple LSTM-based neural networks. We manually define the architecture to get a complex version and a simple version. The architectures are shown in~\cref{tbl:complex_rnn,tbl:simple_rnn}. We use \sgd{} and \dpsgd{} to train these models and show the results in \cref{tbl:rnn-exp}. We observe that in this case also, the simpler architecture outperforms the complex one due to the privacy constraint. Note that the privacy budget is large ($\varepsilon = 7.21$) because the dataset is small.

\begin{table}[!t]
\centering
\caption{Comparing RNN models for \sgd{} and \dpsgd{} ($\varepsilon=7.21$) on \spam{}}
\label{tbl:rnn-exp}
\resizebox{0.9\linewidth}{!}{%
\begin{tabular}{cc|c|c|}
\cline{3-4}
    &  & \multicolumn{2}{c|}{Test accuracy} \\ \hline 
Model   & \# Parameters & \sgd & \dpsgd \\ \hline
Complex (STW) & 496,978               & {\bf 98.80\% }     & 86.00\%            \\ \hline
Simple (PAW) & 50,994                & 98.68\%      & {\bf 93.42\% }       \\ \hline
\end{tabular}
}
\end{table}

\subsection{Regression Experiments}\label{app:exp:regression}
As explained in~\cref{sec:eval:fea:more_features} more features can often be detrimental to private models. To show this we perform some experiments with regression models. We use the \medical{} dataset~\cite{choi_2018} that uses basic patient attributes to estimate insurance costs, which we rescaled (i.e., normalized) for training. We trained several regression neural networks composed of 2, 6, or 12, dense layers, with each layer containing the same number of units and ReLU as the activation function (except for the output layer that contains a single unit with linear activation). For each combination of layers and units we trained the model with \sgd{} and \dpsgd{} (for $\varepsilon=0.95$). In all cases, we use mean squared error as loss function and mean absolute error as a metric for evaluation.

\cref{tbl:lrdpsgd,tbl:lrsgd} show the mean absolute error for the varying layers and number of units for \dpsgd{} and \sgd{}, respectively. As expected, models trained with \sgd{} have lower error across the board as an increasing number of units, and as more layers are used (although this relation is not as strong). However, when using \dpsgd{}, additional units and layers end up being detrimental. In particular, we see that for \sgd{} the lowest error is achieved for $12$ layers each with $60$ units (or $6$ layers each with $120$ units), whereas for \dpsgd{} the lowest error is achieved for $6$ layers each with $10$ units. Further, observe that while the mean absolute error is the same (i.e., $0.063$) for $12$ layers $60$ units and $6$ layers $120$ units for \sgd{}, the same configurations for \dpsgd{} yields vastly different errors ($0.136$ versus $0.333$). We believe this is because the latter has significantly more parameters than the former ($73,561$ versus $40,741$).

\begin{table}[!t]
\centering
\caption{Mean absolute error of various neural networks for regression on \medical{} data trained with \dpsgd{} ($\varepsilon = 0.95$).}
\label{tbl:lrdpsgd}
\small
\begin{tabular}{|l|l|l|l|l|} 
\hline
DPSGD & \multicolumn{4}{c|}{Neurons per Layer} \\ \hline
Layers & 10 & 30 & 60 & 120 \\ \hline \hline
2 & 0.123 & 0.124 & 0.132 & 0.151 \\ \hline
6 & 0.098 & 0.109 & 0.133 & 0.333 \\ \hline
12 & 0.139 & 0.103 & 0.136 & 0.721 \\ \hline
\end{tabular}
\end{table}
\begin{table}[!t]
\centering
\caption{Mean absolute error of various neural networks for regression on \medical{} data trained with \sgd.}
\label{tbl:lrsgd}
\small
\begin{tabular}{|l|l|l|l|l|} 
\hline
SGD & \multicolumn{4}{c|}{Neurons per Layer} \\ \hline
Layers & 10 & 30 & 60 & 120 \\ \hline \hline
2 & 0.08 & 0.074 & 0.07 & 0.068 \\ \hline
6 & 0.074 & 0.068 & 0.064 & 0.063 \\ \hline
12 & 0.154 & 0.073 & 0.063 & 0.067 \\ \hline
\end{tabular}
\end{table}
\subsection{Complexity: width or depth?}\label{app:exp:complexity}
In our experiments, we observed that the cost of achieving privacy grows with the number of parameters of the neural network. However, it is natural to ask if the network's width and depth are factors.

\begin{figure*}[!t]
\centering
\begin{subfigure}{.5\textwidth}
  \centering
  \includegraphics[width=.785\linewidth]{figs/complex_neurons3.pdf}
  \caption{Increasing width (i.e., number of units in each hidden layer).}
  \label{fig:complex_width}
\end{subfigure}%
\begin{subfigure}{.5\textwidth}
  \centering
  \includegraphics[width=.9\linewidth]{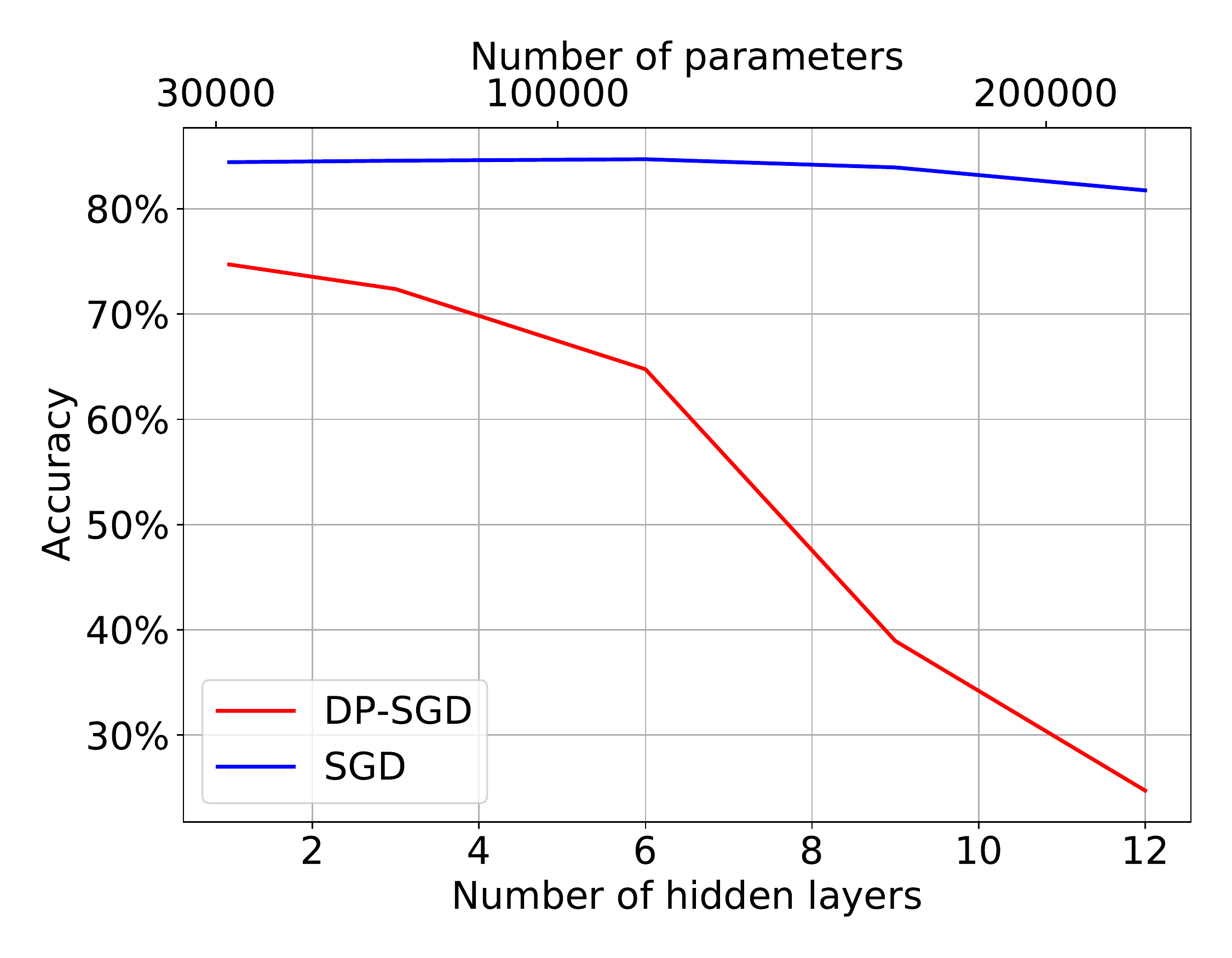}
  \caption{Increasing depth (i.e., number of hidden layers).}
  \label{fig:complex_depth}
\end{subfigure}
\caption{Impact of neural network architecture depth and width on accuracy with \sgd{} and \dpsgd{} on the \svhn{} dataset. Once the number of parameters increases beyond a certain point, the accuracy with \dpsgd{} starts to drop. We observe however, that this drop is relatively modest when increasing width but is drastic when increasing depth. This suggests that architectures with few hidden layers are best suited to \dpsgd.}
\label{fig:complex-width-depth}
\end{figure*}

For example, observe that the PAW CNN model for \cifar{} (\cref{table:cnn_pa_cifar10}) has more layers than the STW CNN model for \cifar{} (\cref{table:cnn_st_cifar10}). But, the latter has more convolutional filters and more total parameters. As a result, the STW model achieves a higher accuracy when trained with \sgd. But, it performs worse than the PAW model when trained with \dpsgd{}~(\cref{tbl:arch-gensearch}). The same situation occurs for the FCN models for \purchase{} and FCN models for \cifar.
To evaluate this systematically, we build a set of models by increasing the number of hidden units (from 4 to 2048) for a single FCN and another set of models by increasing the number of hidden layers (from 1 to 12) but keeping the number of hidden units constant (i.e. 128).
Experiments on \svhn{}~(\cref{fig:complex-width-depth}) show that although increasing model complexity helps achieve better \sgd{} performance, it has a negative impact on the accuracy with \dpsgd{}. For width, increased model complexity increases model performance with \dpsgd{} when the model has a small number of the total number of parameters (e.g., less than $10^4$). However, when the model is already complex, increasing width causes the accuracy to drop further. For depth, the performance of the model trained with \dpsgd{} starts to decrease as we add more layers while this does not occur when the model is trained with \sgd{}. These results suggest that {\bf it is better to use fewer hidden layers.} Indeed, when increasing the number of hidden layers, model performance drops quickly.

\subsection{One-Hot Encoding}\label{app:exp:fs-onehot}

When dealing with categorical features, a ML practitioner will often use one-hot encoding as part of the workflow/pipeline. What is the impact of using one-hot encoding given differential privacy? Surprisingly, even steps of a typical ML workflow such as one-hot encoding need to be reconsidered for differential privacy. To show this, we use the \adult{} dataset which contains 14 attributes, a mix of categorical and numerical attributes. If we one-hot encode the categorical features, we end up with 108 total features. Suppose that instead of one-hot encoding those features, we drop them. In this case, we end up with only 7 features (we drop feature ``fnlwgt'' but also use ``sex'' as a binary feature).

To understand the impact of these given privacy constraints, we train logistic regression models with objective perturbation with one-hot encoding and without. The results are shown in~\cref{tbl:without_one_hot}, where we see that {\bf when $\varepsilon$ is smaller than 0.01, it is better using the dataset without one-hot encoding} even though one-hot encoding clearly helps in obtaining high non-private accuracy. In this case, the crossover epsilon lies in the range $[0.01, 0.05]$.

\begin{table}[!t]
\centering
\caption{Test accuracy with and without one-hot encoding categorical features on the \adult{} data. In the without case, we simply drop the categorical features. The (non private) accuracy without one-hot encoding is 84.44\% and with one-hot encoding it is 84.85\%.}
\label{tbl:without_one_hot}
\resizebox{0.8\linewidth}{!}{%
\begin{tabular}{|c|c|c|}
\hline
$\varepsilon$   & Without One-Hot & With One-Hot \\ \hline \hline
0.1    & 65.93\%±5.36\%              & \textbf{69.65\%±3.33\%}  \\ \hline
0.05   & 65.5\%±4.73\%               & \textbf{66.13\%±4.48\%}  \\ \hline
0.01   & \textbf{64.33\%±4.47\%}     & 59.03\%±9.81\%           \\ \hline
0.001  & \textbf{57.86\%±16.64\%}    & 50.98\%±16.06\%          \\ \hline
0.0001 & \textbf{50.58\%±23.98\%}    & 50.07\%±17.26\%          \\ \hline
\end{tabular}
}
\end{table}

\subsection{\dpsgd{} versus \sgd{}}\label{app:clip_gradients}
We observed several notable differences in behavior when training models with \dpsgd{} and \sgd{}. The first notable difference is that training with \dpsgd{} takes significantly longer than with \sgd{}. In some sense this expected and noted by prior work. However, the difference (at least with its implementation in tensorflow-privacy) is significant: training with \dpsgd{} takes at least an order of magnitude longer than training with \sgd. When training various models on \svhn{} and \spam{} datasets, we calculated the time per epoch in seconds for both \sgd{} and \dpsgd{}. The results are shown in~\cref{tbl:time}, where we see that \dpsgd{} is between 19 and 37 times slower than \sgd{} per epoch. As a result, in our experiments searching for architectures, time spent training models with \dpsgd{} dominates all other operations.

The second is that models trained with \dpsgd{} often outperform \sgd{} whenever the privacy budget is very large. For example, in~\cref{tbl:arch-gensearch}, the \sgd{} accuracy of the PAW model is $45.36\%$ but we see in~\cref{fig:coeps-multiple} that when privacy budget reaches $10^7$ the \dpsgd{} accuracy of the PAW model is about $50.00\%$. After investigating this, we believe this phenomenon is due to the gradient clipping performed by \dpsgd{}. Indeed, when we train the model with \sgd{} but also clip gradients we obtain similar performance.

Furthermore, we found that \dpsgd{} often requires larger learning rates (e.g., 1 or 2) to train models to reasonable accuracies when $\varepsilon$ is large. We also believe this is due to gradient clipping.
\begin{table}[!t]
\centering
\caption{Training time per epoch (in seconds) for \sgd{} and \dpsgd{}. \dpsgd{} is between 19 and 33 times slower.}
\vspace{-6pt}
\label{tbl:time}
\resizebox{0.825\linewidth}{!}{%
\begin{tabular}{ccc|c|c|c|}
\cline{4-5}
    &  &  & \multicolumn{2}{c|}{Time per epoch (s)} \\ \hline 
Data & Model & Parameters & \dpsgd{}  & \sgd{} \\ \hline
\multirow{4}{*}{SVHN} & \multirow{2}{*}{CNN} & 1,817,930 & 23.94 & 1.26  \\ \cline{3-5} 
                      &                      & 600,522  & 20.58 & 0.84  \\ \cline{2-5} 
                      & \multirow{2}{*}{FCN} & 221,066  & 15.96 & 0.50 \\ \cline{3-5} 
                      &                      & 39,818   & 12.60 & 0.34 \\ \hline
\multirow{2}{*}{SPAM} & \multirow{2}{*}{RNN} & 496,978  & 18.80 & 0.56 \\ \cline{3-5} 
                      &                      & 50,994   & 15.28 & 0.42 \\ \hline
\end{tabular}
}
\end{table}
\begin{table}[!tbh]
\centering
\caption{Feature selection with DP-SGD or SGD for Breast Cancer dataset.}
\label{tbl:cfs_sgd_breast}
\resizebox{\linewidth}{!}{%
\begin{tabular}{cc|c|c|c|c|}
\cline{3-6}
                                     &                & \multicolumn{2}{c|}{Selected features} & \multicolumn{2}{c|}{All features} \\ \hline
\multicolumn{1}{|c|}{$\varepsilon$} & \# of Features & SGD                & DP-SGD            & SGD            & DP-SGD           \\ \hline
\multicolumn{1}{|c|}{\multirow{5}{*}{0.0196}}  & 2 & 75.91\% & 75.90\% & \multirow{10}{*}{85.21\%} & \multirow{5}{*}{77.76\%} \\ \cline{2-4}
\multicolumn{1}{|c|}{}               & 28             & 81.81\%            & 80.75\%           &                &                  \\ \cline{2-4}
\multicolumn{1}{|c|}{}               & 48             & 83.33\%            & 80.81\%           &                &                  \\ \cline{2-4}
\multicolumn{1}{|c|}{}               & 68             & 85.30\%            & 79.24\%           &                &                  \\ \cline{2-4}
\multicolumn{1}{|c|}{}               & 88             & 84.91\%            & 78.83\%           &                &                  \\ \cline{1-4} \cline{6-6} 
\multicolumn{1}{|c|}{\multirow{5}{*}{0.00817}} & 2 & 76.25\% & 76.25\% &                           & \multirow{5}{*}{71.10\%} \\ \cline{2-4}
\multicolumn{1}{|c|}{}               & 28             & 81.52\%            & 77.10\%           &                &                  \\ \cline{2-4}
\multicolumn{1}{|c|}{}               & 48             & 84.03\%            & 77.02\%           &                &                  \\ \cline{2-4}
\multicolumn{1}{|c|}{}               & 68             & 84.39\%            & 78.59\%           &                &                  \\ \cline{2-4}
\multicolumn{1}{|c|}{}               & 88             & 85.19\%            & 74.44\%           &                &                  \\ \hline
\end{tabular}
}
\end{table}
\section{Neural Network Architectures}\label{app:arch}
For completeness and reproducibility we present here (some of) the architecture used for experiments in~\cref{sec:eval:arch}. See \cref{table:cnn_pa_cifar10,table:cnn_st_cifar10,table:fcn_pa_purchase,table:fcn_st_purchase,table:fcn_pa_cifar10,table:fcn_st_cifar10,tbl:lenet5-arch,tbl:paw_cnn_mnist,tbl:complex_rnn,tbl:simple_rnn}.
Model architectures used in \randomweights{} experiments are shown in~\cref{sec:exp:factorsflowchat:training_fact}. See~\cref{table:fcn_random}, \cref{tbl:cnn_random} and \cref{tbl:rnn_random}.
\begin{table}[tbh!]
\centering
\caption{CNN architecture for privacy-aware workflow on CIFAR-10.}
\label{table:cnn_pa_cifar10}
\small
\begin{tabular}{|c|c|c|}
\hline
Layer Type     & Output Shape   & Number of Parameters   \\ \hline
MaxPooling     & (16,16,3)      & 0                      \\ \hline
Convolution    & (12,12,16)     & 1216                   \\ \hline
MaxPooling     & (6,6,16)       & 0                      \\ \hline
Convolution    & (6,6,32)       & 4640                   \\ \hline
MaxPooling     & (3,3,32)       & 0                      \\ \hline
Convolution    & (3,3,48)       & 13872                  \\ \hline
Flatten        & (432,)         & 0                      \\ \hline
Dense          & (512,)         & 221696                 \\ \hline
Dropout        & (512,)         & 0                      \\ \hline
Dense          & (10,)          & 5130                   \\ \hline
\multicolumn{3}{|c|}{Total number of parameters: 246554} \\ \hline
\end{tabular}
\end{table}
\begin{table}[tbh!]
\centering
\caption{CNN architecture for standard workflow on CIFAR-10}
\label{table:cnn_st_cifar10}
\small
\begin{tabular}{|c|c|c|}
\hline
Layer Type     & Output Shape    & Number of Parameters   \\ \hline
MaxPooling     & (16,16,3)       & 0                      \\ \hline
Convolution    & (12,12,48)      & 3648                   \\ \hline
MaxPooling     & (6,6,48)        & 0                      \\ \hline
Convolution    & (6,6,96)        & 41568                  \\ \hline
Flatten        & (3456,)         & 0                      \\ \hline
Dense          & (512,)          & 1769984                \\ \hline
Dropout        & (512,)          & 0                      \\ \hline
Dense          & (10,)           & 5130                   \\ \hline
\multicolumn{3}{|c|}{Total number of parameters: 1820330} \\ \hline
\end{tabular}
\end{table}
\begin{table}[tbh!]
\centering
\caption{FCN architecture for privacy-aware workflow on \purchase{}.}
\label{table:fcn_pa_purchase}
\small
\begin{tabular}{|c|c|c|}
\hline
Layer Type  & Number of Neurons  & Number of Parameters \\ \hline
Dense       & 128                & 76928                \\ \hline
Dropout     & 0                  & 0                    \\ \hline
Dense       & 100                & 12900                \\ \hline
\multicolumn{3}{|c|}{Total number of parameters: 89828} \\ \hline
\end{tabular}
\end{table}
\begin{table}[tbh!]
\centering
\caption{FCN architecture for standard workflow on \purchase{}.}
\label{table:fcn_st_purchase}
\small
\begin{tabular}{|c|c|c|}
\hline
Layer Type  & Number of Neurons  & Number of Parameters \\ \hline
Dense               & 1024                       & 615424                        \\ \hline
Dropout             & 0                          & 0                             \\ \hline
Dense               & 256                        & 262400                        \\ \hline
Dropout             & 0                          & 0                             \\ \hline
Dense               & 100                        & 25700                         \\ \hline
\multicolumn{3}{|c|}{Total number of parameters: 903524}                \\ \hline
\end{tabular}
\end{table}
\begin{table}[tbh!]
\centering
\caption{FCN architecture for privacy-aware workflow on \cifar.}
\label{table:fcn_pa_cifar10}
\small
\begin{tabular}{|c|c|c|}
\hline
Layer Type  & Number of Neurons  & Number of Parameters  \\ \hline
Dense               & 64                         & 19264                         \\ \hline
Dropout             & 0                          & 0                             \\ \hline
Dense               & 10                         & 650                           \\ \hline
\multicolumn{3}{|c|}{Total number of parameters: 19914}                 \\ \hline
\end{tabular}
\end{table}
\begin{table}[tbh!]
\centering
\caption{FCN architecture for standard workflow on \cifar.}
\label{table:fcn_st_cifar10}
\small

\begin{tabular}{|c|c|c|}
\hline
Layer Type  & Number of Neurons  & Number of Parameters  \\ \hline
Dense               & 2048                       & 616448                        \\ \hline
Dropout             & 0                          & 0                             \\ \hline
Dense               & 10                         & 20490                         \\ \hline
\multicolumn{3}{|c|}{Total number of parameters: 636938}                \\ \hline
\end{tabular}
\end{table}
\begin{table}[tbh!]
\centering
\caption{Architecture of LeNet-5 (for \mnist).}
\label{tbl:lenet5-arch}
\small
\begin{tabular}{|c|c|c|}
\hline
Layer Type        & Output shape   & \# of parameters   \\ \hline
Convolution       & (28,28,16)     & 156                \\ \hline
AveragePooling    & (14,14,6)      & 0                  \\ \hline
Convolution       & (10,10,16)     & 2416               \\ \hline
AveragePooling    & (5,5,16)       & 0                  \\ \hline
Flatten           & (400,)         & 0                  \\ \hline
Dense             & (120,)         & 48120              \\ \hline
Dense             & (84,)          & 10164              \\ \hline
Dense             & (10,)          & 850                \\ \hline
\multicolumn{3}{|c|}{Total number of parameters: 61706} \\ \hline
\end{tabular}
\end{table}
\begin{table}[tbh!]
\centering
\caption{Architecture of PAW model for \mnist{}.}
\label{tbl:paw_cnn_mnist}
\small
\begin{tabular}{|c|c|c|}
\hline
Layer Type      & Output shape    & \# of parameters    \\ \hline
MaxPooling      & (14,14,1)       & 0                   \\ \hline
Convolution     & (10,10,16)      & 416                 \\ \hline
MaxPooling      & (5,5,16)        & 0                   \\ \hline
Convolution     & (5,5,32)        & 4640                \\ \hline
Flatten         & (800,)          & 0                   \\ \hline
Dense           & (64,)           & 51264               \\ \hline
Dropout         & (64,)           & 0                   \\ \hline
Dense           & (10,)           & 650                 \\ \hline
\multicolumn{3}{|c|}{Total number of parameters: 56970} \\ \hline
\end{tabular}
\end{table}
\begin{table}[tbh!]
\centering
\caption{Architecture of complex RNN for \spam.}
\label{tbl:complex_rnn}
\small
\begin{tabular}{|c|c|c|}
\hline
Layer Type  & Number of Neurons  & Number of Parameters  \\ \hline
Embedding   & 150                & 50000                 \\ \hline
LSTM        & 256                & 314368                \\ \hline
Dense       & 512                & 131584                \\ \hline
Dropout     & 0                  & 0                     \\ \hline
Dense       & 2                  & 1026                  \\ \hline
\multicolumn{3}{|c|}{Total number of parameters: 496978} \\ \hline
\end{tabular}
\end{table}
\begin{table}[tbh!]
\centering
\caption{Architecture of simple RNN for \spam.}
\label{tbl:simple_rnn}
\small

\begin{tabular}{|c|c|c|}
\hline
Layer Type  & Number of Neurons  & Number of Parameters \\ \hline
Embedding   & 150                & 50000                \\ \hline
LSTM        & 4                  & 880                  \\ \hline
Dense       & 16                 & 80                   \\ \hline
Dropout     & 0                  & 0                    \\ \hline
Dense       & 2                  & 34                   \\ \hline
\multicolumn{3}{|c|}{Total number of parameters: 50994} \\ \hline
\end{tabular}
\end{table}
\begin{table}[tbh!]
\centering
\caption{FCN architecture for \randomweights{} on \mnist{}.}
\label{table:fcn_random}
\small
\begin{tabular}{|c|c|c|}
\hline
Layer Type  & Number of Neurons  & Number of Parameters \\ \hline
Dense               & 1024                       & 803840                        \\ \hline
Dense               & 512                        & 524800                           \\ \hline
Dense               & 64                        & 32832                       \\ \hline
Dense               & 10                        & 650                        \\ \hline
\multicolumn{3}{|c|}{Total number of parameters: 1,362,122 (Trainable: 33,482)}                \\ \hline
\end{tabular}
\end{table}
\begin{table}[tbh!]
\centering
\caption{CNN architecture for \randomweights{} on \mnist{}.}
\label{tbl:cnn_random}
\small
\begin{tabular}{|c|c|c|}
\hline
Layer Type      & Output shape    & \# of parameters    \\ \hline
Convolution     & (24,24,32)      & 832                 \\ \hline
MaxPooling      & (12,12,32)       & 0                   \\ \hline
Convolution     & (10,10,64)      & 18496                 \\ \hline
MaxPooling      & (5,5,64)        & 0                   \\ \hline
Flatten         & (1600,)          & 0                   \\ \hline
Dense           & (64,)           & 102464               \\ \hline
Dense           & (10,)           & 650                 \\ \hline
\multicolumn{3}{|c|}{Total number of parameters: 122,442 (Trainable: 19,328)} \\ \hline
\end{tabular}
\end{table}
\begin{table}[tbh!]
\centering
\caption{RNN architecture for \randomweights{} on \spam{}.}
\label{tbl:rnn_random}
\small
\begin{tabular}{|c|c|c|}
\hline
Layer Type  & Number of Neurons  & Number of Parameters  \\ \hline
Embedding   & 150                & 50,000                 \\ \hline
LSTM        & 256                & 314,368                \\ \hline
Dense       & 512                & 131,584                \\ \hline
Dense       & 2                  & 1,026                  \\ \hline
\multicolumn{3}{|c|}{Total number of parameters: 496,978 (Trainable: 132,610)} \\ \hline
\end{tabular}
\end{table}

\end{document}